\documentclass[final,5p,times,twocolumn]{elsarticle} 

\usepackage{graphicx}

\usepackage{amssymb}

\usepackage[colorlinks=true,linkcolor=blue]{hyperref}

\journal{Nuclear Instruments and Methods A}

\begin{document}

\begin{frontmatter}

\title{Silicon Photomultipliers in Particle and Nuclear Physics}

\author{Frank Simon}
\ead{fsimon@mpp.mpg.de}
\address{Max-Planck-Institut f\"ur Physik, F\"ohringer Ring 6, 80805 M\"unchen, Germany}

\begin{abstract}
Following first large-scale applications in highly granular calorimeters and in neutrino detectors, Silicon Photomultipliers have seen a wide adoption in accelerator-based particle and nuclear physics experiments. Today, they  are used for a wide range of different particle detector types, ranging from calorimeters and trackers to particle identification and veto detectors, large volume detectors for neutrino physics and timing systems. This article reviews the current state and expected evolution of these applications, highlighting strengths and limitation of SiPMs and the corresponding design choices in the respective contexts. General trends and adopted technical solutions in the applications are discussed. 
\end{abstract}


\end{frontmatter}

\tableofcontents

\section{Introduction}
\label{sec:Intro}

Silicon Photomultipliers\footnote{In this paper, the term ``Silicon Photomultiplier'' is used generically to refer to silicon-based multipixel Geiger-mode-operated avalanche photodiodes, independent of producer and concrete design. A wide variety of names and acronyms, used by different producers and R\&D groups exists. These will only be used here to refer to  specific variants of SiPMs.} (SiPMs), are seeing an increasingly wide-spread use in accelerator-based particle and nuclear physics experiments. These photon detectors have a number of features that make them very attractive for collider and fixed-target experiments, such as a very small size, a high photon detection efficiency, insensitivity to magnetic fields and to excessively large signals from ionizing particles, high intrinsic gain resulting in simplifications in the readout electronics, a fast response, a level of radiation hardness sufficient for many practical applications, and relatively low operating voltages. On the other hand, limitations such as the finite dynamic range imposed by the number of micro-cells in the device, the voltage-dependent excess noise factor, and the strong temperature dependence of the breakdown voltage have to be taken into account in the detector design. These and other strengths, features and limitations are reviewed in more detail in other articles in this volume. In recent years, commercially available devices have reached a high level of uniformity, favorable performance parameters and a price point that enables large-scale use in a variety of different contexts. 

The most common application of SiPMs in accelerator-based experiments is the readout of scintillation light from organic, inorganic and cryogenic scintillators. In addition, there is active development and prototyping in the area of Cherenkov detectors. In the past 15 years, the use of SiPMs has evolved from first large-scale applications in a sampling calorimeter with plastic scintillator active elements with approximately 8000 sensors to a wide range of different calorimeter applications, tracking detectors, particle identification and veto systems, photon detection systems in large-volume detectors as well as developments towards timing detectors for measurements at the few 10 ps level. 

This article will give a broad overview over applications of SiPMs in particle and nuclear physics experiments at accelerators, also highlighting the evolution from first applications in the mid 2000's to modern systems today (mid 2018). These are grouped according to the main purpose of the systems, although the assignment sometimes is not unique, for example in the area of veto and timing systems. The focus is on applications in physics experiments and larger R\&D projects, rather than small development activities. Given the large number of projects using SiPMs today, only a selection of examples can be shown, reflecting the personal bias of the author. Finally, a summary of the main trends in the adopted technical solutions is presented. 

\section{Pioneering experiments: CALICE \& T2K}
\label{sec:Pioneers}

The first detector with a large number of SiPMs operated in a particle beam from 2006 on was the CALICE Analog Hadron Calorimeter (AHCAL) physics prototype \cite{collaboration:2010hb}, with 7,608 photon sensors, following  the construction and test of a proof-of-principle prototype in 2004 \cite{Andreev:2004uy}. The AHCAL went through an extensive test beam program at DESY, CERN and Fermilab, taking data until 2011. From 2007 on, the AHCAL was complemented by a Tail Catcher and Muon Tracker (TCMT) \cite{CALICE:2012aa}, also making use of SiPMs. The first particle physics experiment to adopt SiPMs at a large scale was T2K \cite{Abe:2011ks}, with several subsystems of the near detector ND280 and the on-axis near detector INGRID \cite{Abe:2011xv} based on SiPM readout. In total, 56,000 photon sensors are used in the experiment. Following an extensive R\&D program \cite{Yokoyama:2010qa}, which started with the selection of SiPMs as the photon detectors for the near detector complex at the end of 2005, the full detector saw the first beam in 2010. 

Both CALICE and T2K use the SiPMs to read out scintillation light from plastic scintillators. Wavelength-shifting fibers embedded in the plastic scintillator elements are used to collect the scintillation light, guide it to the photon sensors, and match the wavelength to the sensitivity in the green spectral region of the first-generation SiPMs used in these experiments. The SiPM signals are brought to ASIC-based front-end electronics via coaxial cables, which also supply the bias voltage of the photon sensor. 

\begin{figure}
\centering
\includegraphics[width = 0.7\columnwidth]{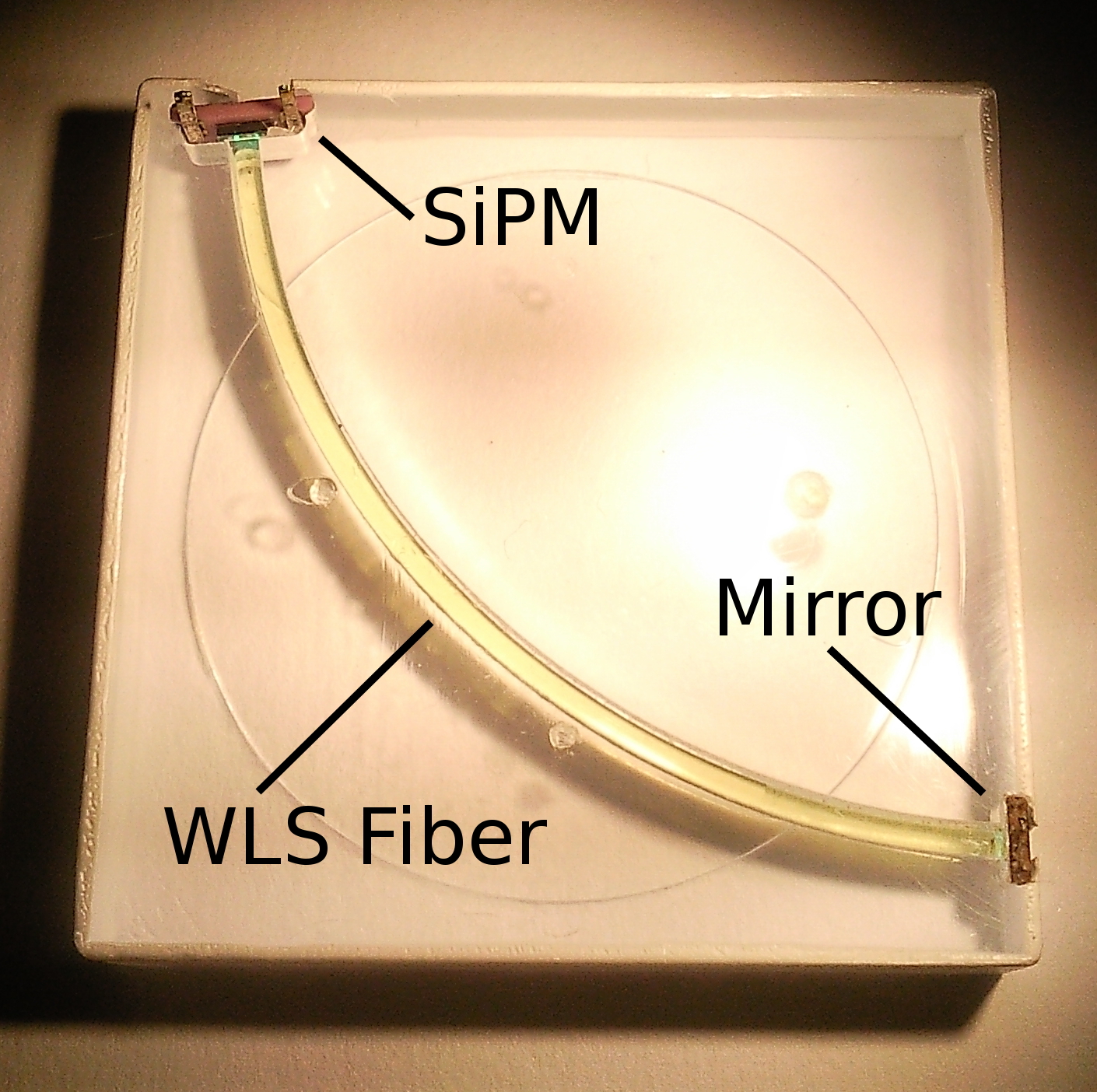}
\caption{Photograph of a $3 \times 3~\mathrm{cm}^2$, 5~mm thick scintillator tile with embedded wavelength-shifting fiber coupled to a SiPM integrated into the tile used for the CALICE analog hadron calorimeter. Figure taken from \cite{Simon:2010hf}\label{fig:Pioneers:CALICETile}}
\end{figure}

Figure \ref{fig:Pioneers:CALICETile} shows the basic detection element of the CALICE AHCAL physics prototype, a  $3 \times 3 \times 0.5~$cm$^2$ plastic scintillator tile with an embedded wavelength-shifting fiber with a diameter of 1 mm coupled to a SiPM. To reduce the number of channels of the prototype, larger tiles with a size of $6 \times 6$ and $12 \times 12~\mathrm{cm}^2$ are used in the outer regions and rear part of the detector. The SiPMs for the prototype have an active area of $1.1 \times 1.1~\mathrm{mm}^2$, with 1156 pixels with a size of $32 \times 32~\mu\mathrm{m}^2$, produced by MEPhI/CPTA. The coupling of the fiber in the tile to the SiPM is via a 50 to 100~$\mu$m wide air gap, with the lateral positioning of the SiPM provided by a milled groove. 

The inherent limitation of the dynamic range of the SiPMs due to the finite number of pixels results in a saturation behavior which is particularly critical in the core of electromagnetic showers. This saturation effect is addressed by the cell-by-cell application of a correction algorithm which gives the expected true number of photons for a given number of fired SiPM pixels, as discussed in detail in \cite{collaboration:2010rq}, resulting in the recovery of a linear response of the calorimeter over a wide energy range. While the uncertainty of this saturation correction is the leading systematic in the energy reconstruction for electromagnetic showers of 40~GeV and above \cite{collaboration:2010rq}, it was found to be negligible for hadronic showers at all energies \cite{Adloff:2012gv}. 
With local software compensation techniques used for energy reconstruction, the prototype achieved a hadronic energy resolution of $\sim44\%/\sqrt{E \mathrm{[GeV]}} \oplus 1.8\%$ \cite{Adloff:2012gv}, with a response within 1.5\% of linearity. 
Following the successful tests of the CALICE AHCAL physics prototype, the technology for highly granular scintillator-based calorimeters was further developed, as discussed in \autoref{ssec:HighGranularity}.

\begin{figure}
\centering
\includegraphics[width = 0.4\columnwidth]{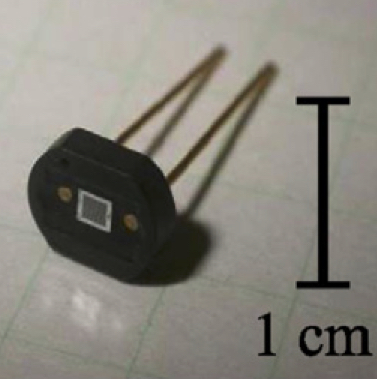}
\hspace{0.025\columnwidth}
\includegraphics[width = 0.515\columnwidth]{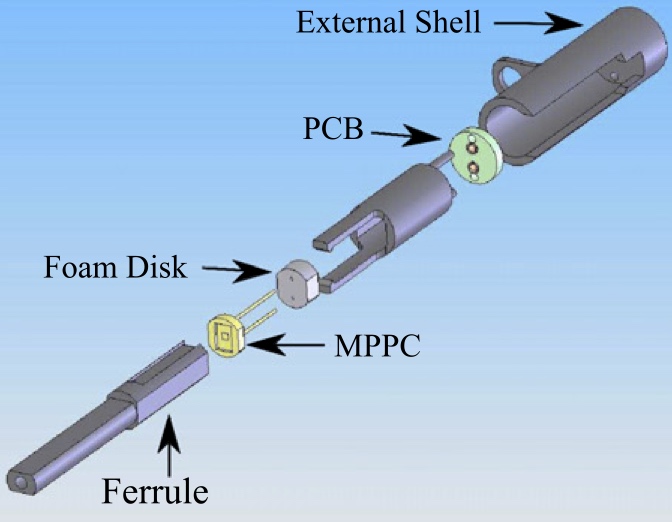}
\caption{Left: Photograph of a T2K SiPM in ceramic package, taken from \cite{Abe:2011ks}. Right: Exploded view of the connector used to connect a clear optical fiber to a SiPM in the T2K near detector P0D, taken from \cite{Assylbekov:2011sh}\label{fig:Pioneers:T2KConnector}}
\end{figure}

In T2K, SiPMs are used in electromagnetic calorimeters, scintillator-based tracking systems and muon detectors in the near detector system, demonstrating their versatility. One common type of SiPM with an active area of $1.3 \times 1.3~\mathrm{mm}^2$ and 667 pixels, produced by Hamamatsu specifically for the experiment, is used for all systems in T2K. The left panel of  \autoref{fig:Pioneers:T2KConnector} shows the SiPM used by T2K. Different schemes for coupling  the wavelength-shifting fiber and the photon sensor are used, depending on the layout of the subdetector. Here, the photon sensors are not embedded in the scintillator material, but are coupled to wavelength-shifting fibers embedded in the scintillator to the sensors by means of a dedicated connector that has been developed to ensure an optimal coupling. Several variations of this connector are in use in the experiment, with \autoref{fig:Pioneers:T2KConnector} (right) showing the one adopted for the  $\pi^0$ detector P0D. The SiPM signal is carried to the front-end electronics via a coaxial cable, which is also used to supply the bias voltage for the sensor. 

The T2K systems using SiPMs are the INGRID on-axis detector \cite{Abe:2011xv}, and the $\pi^0$ detector P0D \cite{Assylbekov:2011sh}, the fine-grained tracker FGD \cite{Amaudruz:2012agx}, the electromagnetic calorimeter ECAL \cite{Allan:2013ofa} and the side muon range detector SMRD \cite{Aoki:2012mf} of the  off-axis near detector. The ND280 systems are located in a 0.2~T dipole field, while INGRID sits in an essentially field-free region. INGRID, P0D, ECAL and FGD use different types of extruded scintillator bars with a wavelength-shifting fiber inserted in a central hole, while SMRD uses larger scintillator plates with a meandering fiber. In general the fibers are only read out on one end, with the other end mirrored. The ECAL uses a mix of single- and double-sided readout. The bars read out on both ends show a higher efficiency to minimum-ionizing particles due to the larger collected signal, and the reduced impact of light absorption along the fiber. The 34 layer downstream ECAL, with a thickness of 11~$X_0$, achieved an energy resolution of approximately 10\% and an angular resolution of 4$^\circ$ for normal-incident 1~GeV electrons in beam tests. The time resolution for tracks and showers is better than 1~ns \cite{Allan:2013ofa}.

In both CALICE and T2K, specific measures were taken to cope with the temperature dependence of the breakdown voltage of the SiPMs, which results in changes of the signal amplitudes with changing temperature, visible in diurnal variations of the detector response. In T2K, cooling loops in the front-ends close to the photon sensors reduce but do not eliminate the fluctuations. They are corrected for by making frequent calibration runs to determine the gain, and in the case of CALICE also by applying a temperature correction based on measurements of the temperature within the detector volume. The temperature measurements are used to a apply a correction based on the measured temperature dependence of the gain and photon detection efficiency in the data analysis. 

In these early large-scale applications the sensors have demonstrated a high degree of reliability, with sensor failures on the permille level (CALICE) or below (T2K) in the first years of operation. In both projects, the SiPM-based detectors have in general achieved or exceeded the performance expected during the design phase, underlining the capabilities of SiPMs as photon detectors for accelerator-based particle physics experiments.

\section{Applications in calorimetry}
\label{sec:Calorimetry}

Today, calorimetry is still the dominant area of use of SiPMs in particle physics, ranging from applications in smaller-scale experiments to detector upgrades for the LHC and developments for future collider experiments. In this section, a few representative examples of calorimeter applications will be discussed, covering scintillating-fiber, scintillator-tile, crystal and cryogenic detectors, as well as highly granular ``imaging'' calorimeters based on plastic scintillator elements.

\subsection{Classical scintillator calorimeters}
\label{ssec:ScintCalo}

The compactness, low operation voltage, insensitivity to magnetic fields as well as the high photon detection efficiency make SiPMs an interesting alternative to classical photomultipliers and other light detectors for a wide range of different calorimeter concepts. Here, we consider three examples, a scintillating fiber electromagnetic calorimeter, the upgrade of a plastic scintillator-based hadronic calorimeter at the LHC and a crystal-based electromagnetic calorimeter.

\begin{figure}
\centering
\includegraphics[width = 0.65\columnwidth]{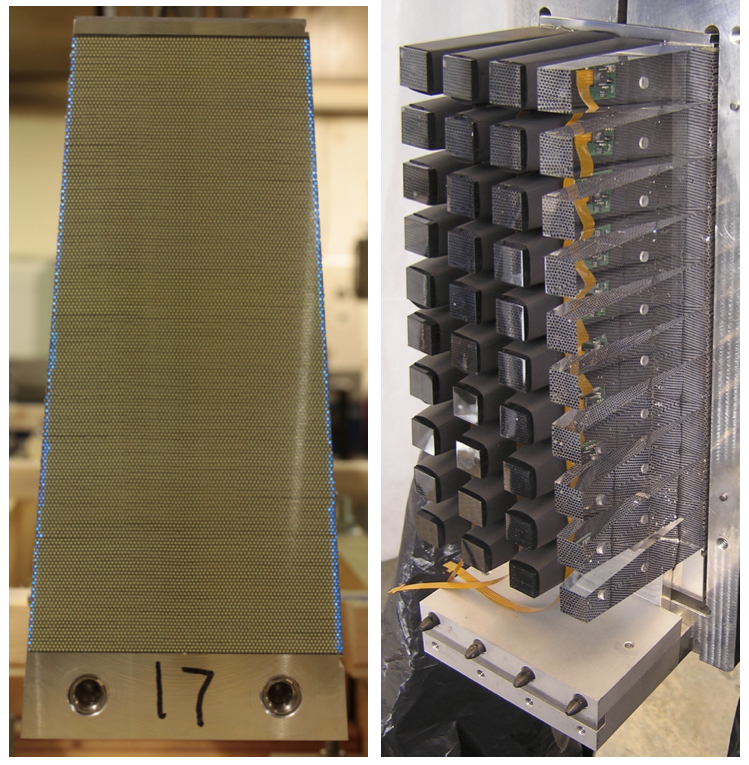}
\caption{One segment of the GlueX calorimeter, showing the matrix of scintillating fibers embedded in lead (left) and the light guides attached to the matrix (right). To the ends of the light guides  $13 \times 13~\mathrm{mm}^2$ 16-channel SiPM arrays are coupled to detect the light. Figure taken from \cite{Beattie:2018xsk}. \label{fig:Calorimetry:GlueX}}
\end{figure}

GlueX at JLab is also among the early adopters of SiPM technology in particle and nuclear physics, installing a SiPM-based calorimeter system in 2013. The barrel electromagnetic calorimeter (BCAL) \cite{Beattie:2018xsk} is used to measure photons in the energy range of 50~MeV to several GeV. It is a sampling calorimeter with scintillating fibers embedded in corrugated sheets of lead. The photons produced in the 1~mm diameter, 390~cm long Kuraray SCSF-78MJ double-clad blue-green emitting fibers which read out on both ends. Light guides with varying size are used to collect the light from the polished fiber ends and guide it to the photon sensors, as illustrated in \autoref{fig:Calorimetry:GlueX}. Each light guide ends in a $4 \times 4$ array of $3 \times 3~\mathrm{mm}^2$ SiPMs with a total size of $13 \times 13~\mathrm{mm}^2$ and a total active area of 144~mm$^2$ produced by Hamamatsu. SiPMs were chosen primarily for their insensitivity to magnetic fields, allowing the placement of the photon sensors directly at the end of the light guides. Each of the SiPMs has 3,600 pixels with a size of $50 \times 50~\mu\mathrm{m}^2$. Technically, the arrays provide the possibility for an independent biasing of each row, and an independent readout of each of the 16 SiPMs. Here, one array is connected to a common bias channel, and all 16 signal channels are combined to provide one common output of the sensor.  In total 3,840 SiPM arrays are installed in the calorimeter. To reduce the number of electronics channels, the analog signals of the photon sensors at larger radius are summed up in a 1:2:3:4 scheme, providing four longitudinal segments with increasing depth.

The GlueX BCAL achieves an energy resolution of 5.2\%/$\sqrt{E[\mathrm{GeV}]}$ with a constant term of 3.6\% and a time resolution of 150~ps for 1~GeV electromagnetic showers. This is comparable to the performance parameters of the KLOE electromagnetic calorimeter \cite{Adinolfi:2002zx}, which is based on scintillating fibers in a lead matrix read out with photomultipliers. The GlueX BCAL is modeled after the KLOE calorimeter, using a very similar sampling structure. The KLOE electromagnetic calorimeter achieved an energy resolution of  5.7\%/$\sqrt{E[\mathrm{GeV}]}$ and a time resolution of 54~ps/$\sqrt{E[\mathrm{GeV}]} \oplus 50~\mathrm{ps}$ after sophisticated corrections. The fact that the BCAL achieves a similar performance demonstrates that already early SiPM-based detectors are as capable as systems using PMTs. 

SiPMs are also used to upgrade existing detectors, such as the hadronic barrel and endcap calorimeters of CMS \cite{CMS:2012tda}. For these two systems, the hybrid photodiodes (HPDs) used in the original detector are in the process of being replaced by SiPMs. The new sensors for the endcap calorimeter are already installed, with the barrel calorimeter to follow in the second long LHC shutdown \cite{Strobbe:2016ojb}. This is the first large-scale use of SiPMs in the high radiation environment of a hadron collider, and follows the replacement of HPDs by SiPMs in the CMS outer hadron calorimeter, where 2,400 SiPMs are already in operation since 2015 \cite{Lobanov:2015jla}. The motivation for this replacement is the need to increase the ability of the detector to separate showers and to identify particle signatures in the deeper regions of the calorimeter which have a lower occupancy, and to cope with increasing radiation damage of the plastic scintillator, which results in a position-dependent reduction of the light yield. In this respect, the SiPMs present two key advantages over the HPDs, namely an approximately three times higher photon detection efficiency which directly reduces the impact of the light yield degradation of the scintillator system, and a much smaller physical size, which allows to install a larger number of photon sensors, providing a finer longitudinal segmentation of the calorimeter. This finer segmentation, which is chosen in a way that provides a uniform energy deposit in each of the readout depths for the averaged shower profile to provide a uniform signal to noise ratio for the different segments, improves the shower separation capabilities of the detector, and enables a more precise monitoring of and correction for the depth-dependent radiation damage using minimum-ionizing particles and other probes. 

\begin{figure}
\centering
\includegraphics[width = 0.75\columnwidth]{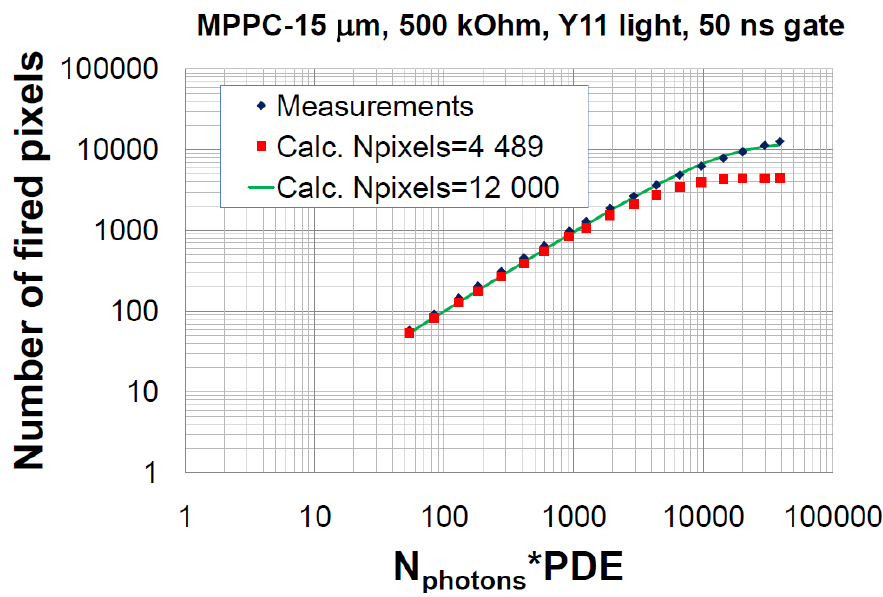}
\caption{Measurement of response of SiPM to a light pulse matching the time distribution of the scintillation light collected via a Y11 wavelength-shifting fiber, compared to the response expected when taking only the number of pixels in the photon sensor of $\sim$4,500, and the expected response when considering the fast recovery time of 4 ns, resulting in an effective number of pixels of  $\sim$12,000. Figure taken from \cite{CMS:2012tda}. \label{fig:Calorimetry:CMS_SiPMLinearity}}
\end{figure}

In total, 16,000 SiPMs manufactured by Hamamatsu will be installed in this so-called phase I upgrade of the CMS main hadron calorimeter. To reach a sufficient dynamic range, SiPMs with 4,500 pixels/mm$^2$, corresponding to a pixel size of $15\times15~\mu\mathrm{m}^2$, have been selected. Two different sensor sizes are used to match the varying number of fibers guiding the light from the scintillator layers. By choosing SiPMs with a short recovery time of less than 10~ns, which allows a pixel to fire more than once during the collection time of the scintillation light of $\sim$ 40 ns, the effective pixel count, and thus the dynamic range, is increase by a factor of approximately three \cite{CMS:2012tda}. This extends the range of a quasi-linear response of the SiPMs, as shown in \autoref{fig:Calorimetry:CMS_SiPMLinearity}, while also resulting in a slight distortion of the pulse due to an accentuation of the late tail of the photon distribution where photon sensor saturation has abated. The effect of these distortions is expected to be negligible.  The short recovery time is achieved by using SiPMs with low quench resistors. The adopted sensors use a quench resistor of 500~k$\Omega$. The temperature dependence of the sensor characteristics is controlled by precise temperature stabilization provided by Peltier elements steered via a software control loop taking the temperature measurement of a sensor close to the SiPM as input. 

This upgrade addresses the needed longevity improvements for the barrel hadron calorimeter for the HL-LHC, removing the need for dedicated phase II upgrades \cite{Collaboration:2283187}. The endcap calorimeter, on the other hand, will be completely replaced by a new system in the third long LHC shutdown, as discussed in \autoref{ssec:HighGranularity}.

\begin{figure}
\centering
\includegraphics[width = 0.455\columnwidth]{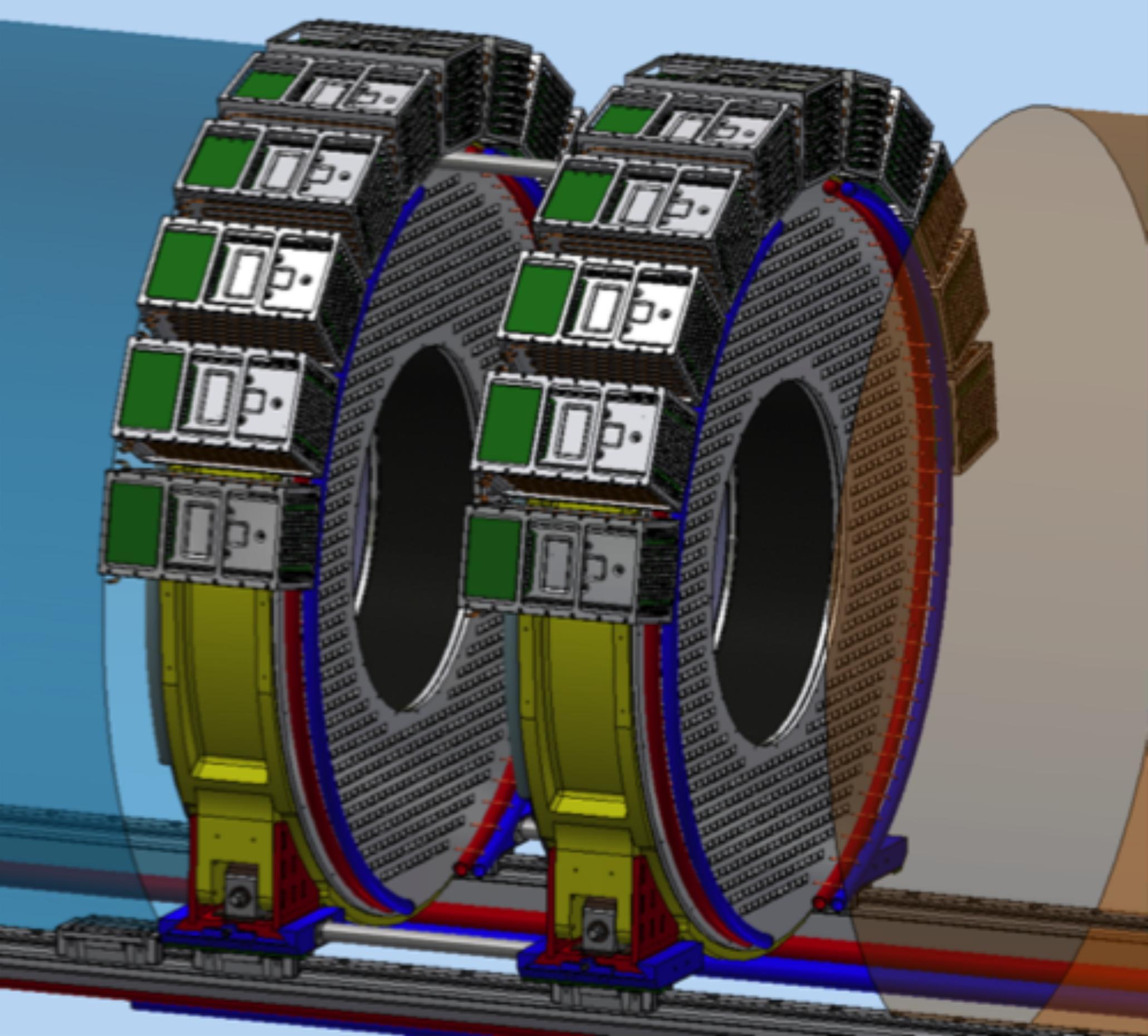}\vspace{1mm}
\includegraphics[width = 0.528\columnwidth]{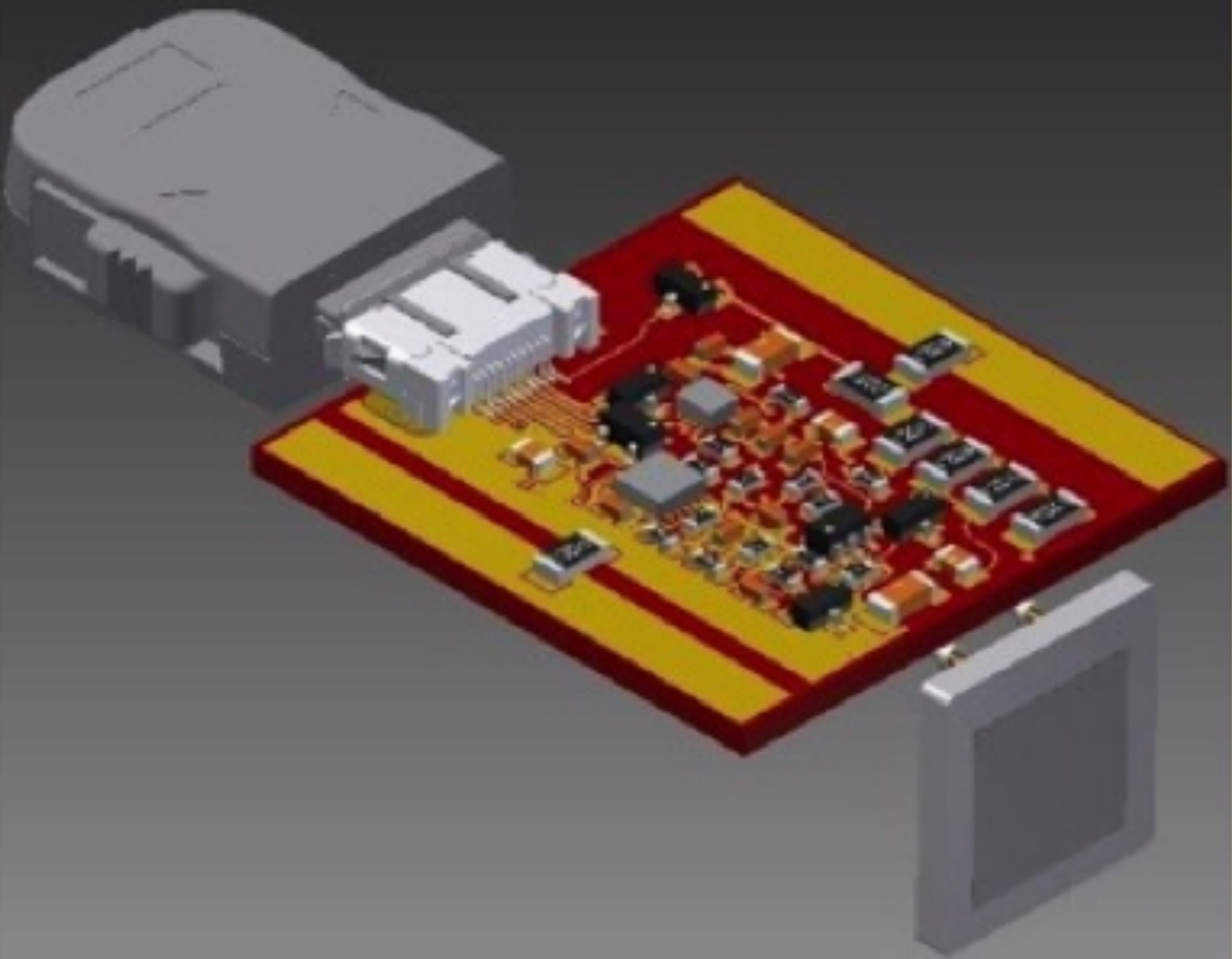}
\caption{Left: Design of the Mu2e ECAL, with two rings of 674 pure CsI crystals, each read out by two SiPM arrays. Right: Rendering of a frontend electronics module with a SiPM array, which is used to read out a single crystal. Figure taken from \cite{Atanov:2016itr} \label{fig:Calorimetry:Mu2eECAL}}
\end{figure}

In addition to the common applications of reading out plastic scintillators, SiPMs are also well-suited for the readout of crystals used in electromagnetic calorimeters. An example for this is the electromagnetic calorimeter of the Mu2e experiment at Fermilab \cite{Atanov:2018bfv}. This detector, currently still in the R\&D phase, uses pure CsI crystals read out by custom-designed arrays of SiPMs. These arrays consist of a matrix of $2\times 3$ electrically separated SiPMs, each with an active area of $6 \times 6~\mathrm{mm}^2$. Two such arrays are used to read out one crystal with a lateral size of $34 \times 34~\mathrm{mm}^2$. In total, the Mu2e electromagnetic calorimeter will consist of 1,348 crystals, arranged in two annuli separated by 75~cm. \autoref{fig:Calorimetry:Mu2eECAL} shows the design of the calorimeter as well as a rendering of a frontend module with attached SiPM array used to read out a single crystal. 

The SiPMs used in Mu2e are special UV-extended sensors to provide sensitivity to the 315~nm photons of the fast scintillation component of CsI. This is crucial to achieve a time resolution of better than 500~ps, as required by the physics goals of the experiment. Test beam measurements using crystals with a lateral size of $30 \times 30~\mathrm{mm}^2$ read out by one $12 \times 12~\mathrm{mm}^2$ Hamamatsu SiPM array ($4\times 4$ sensors) with a silicon entrance window have demonstrated that the requirements of the experiment can be met, with a time resolution of 110~ps and an energy resolution of 7\% (dominated by leakage) at 100~MeV \cite{Atanova:2017ppl}. Here, the time resolution surpasses the one obtained with the same crystals read out by a PMT. A ``Module-0'' demonstrator has been constructed using sensors from AdvanSiD, Hamamatsu and SensL as well as crystals from different manufacturers to evaluate the performance of the different devices \cite{Atanov:2018bfv}. The detector has recently been exposed to particle beams in Frascati, with data analysis ongoing.

\subsection{Cryogenic calorimeters}
\label{ssec:ScintCryo}

\begin{figure}
\centering
\includegraphics[width = 0.75\columnwidth]{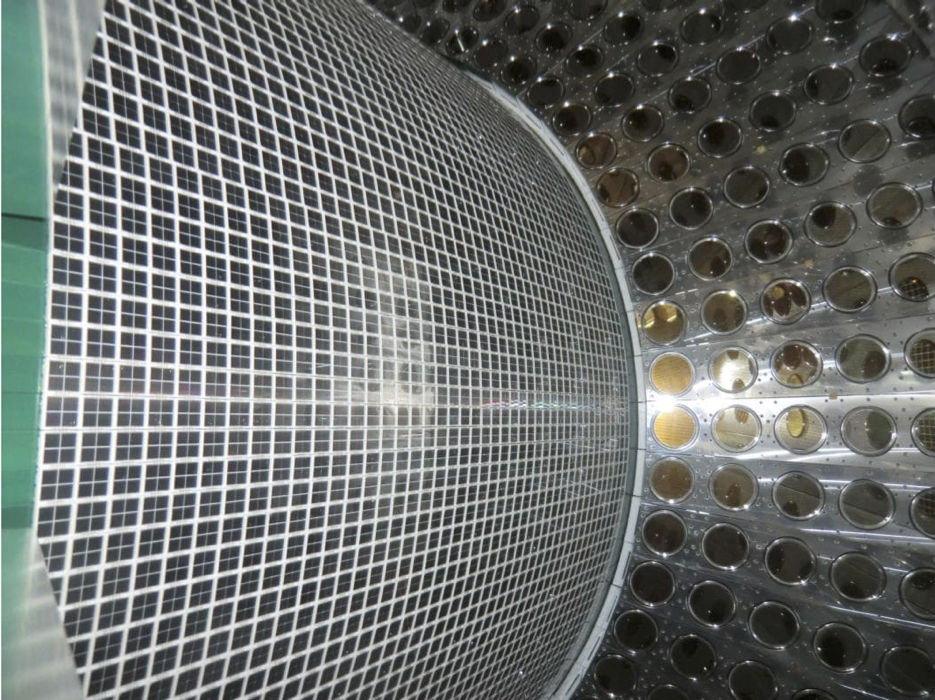}
\caption{Inside of the MEG II liquid Xe calorimeter showing the SiPMs on the inner face of the detector, as well as the PMTs still used on the sides. Figure taken from \cite{Baldini:2018nnn} \label{fig:Calorimetry:MEG_SiPMs}}
\end{figure}

The MEG experiment at PSI, which finished data taking in 2013, was using a completely different calorimeter technology, a liquid Xe calorimeter to measure 52.8~MeV photons from lepton-flavor violating decay $\mu^+ \rightarrow e^+\gamma$. The experiment is currently being upgraded to MEG II  \cite{Baldini:2018nnn}  for an increase in intensity and resolution. In the original liquid Xe calorimeter, the scintillation light created by particle showers in the cryogenic volume was detected by photomultipliers. In the upgraded detector, Hamamatsu-made SiPMs replace the PMTs  on the surface of the calorimeter facing the beam pipe, which is the inner surface of the detector, as shown in \autoref{fig:Calorimetry:MEG_SiPMs}. This increases surface coverage and thus the light collection, and the granularity of the readout, resulting in  better energy and position resolution of the detector. In addition, the SiPMs contribute less material than the PMTs, leading to a 9\% improvement of the efficiency for 50~MeV photons with respect to the original detector. This is due to the reduced interaction probability before the photons enter the liquid Xe volume. The outer and end surfaces retain the PMT readout. 

The key challenge for this detector upgrade is the development of SiPMs which are sensitive to liquid Xe scintillation light ($\lambda \, \sim$ 175~nm).  To achieve this  VUV sensitivity, several modifications with respect to standard SiPMs are implemented in the MEG II sensors \cite{Ootani:2015cia}. The contact layer is thinned to improve the efficiency for the detection of short wavelengths. In addition, the protection layer of resin and the anti-reflective coating is removed to improve the optical matching between the liquid Xe and the sensor surface. The protection of the sensor is instead provided by a VUV-transparent quartz window. Together, these modifications enable the detection of VUV photons, while unmodified devices provide no sensitivity in the relevant spectral range. The devices use a metal quench resistor to achieve a smaller temperature coefficient than when using polysilicon. In addition, the sensors use up-to-date crosstalk suppression, limiting optical crosstalk to 5\% at nominal operation conditions. A photon detection efficiency of $\sim$20\% is achieved for liquid Xe scintillation light \cite{Ootani:2015cia,Ogawa:2017xof}, at least on a par with the one provided by the previously used PMTs. 

The MEG II calorimeter uses 4,092 discrete arrays of four $6\times 6~\mathrm{mm}^2$ SiPMs with a pixel size of $50 \times 50$~$\mu$m connected in series. This arrangement is chosen to achieve a large detection surface per channel, while at the same time limiting the capacitance of the sensor. The series connection of the individual sensors enables a better time resolution than other connection schemes by preserving a short signal rise time, as discussed in \autoref{sec:Timing}. The upgraded calorimeter is expected to reach a time resolution of 50~ps or better for signal photons with an energy of 52.8~MeV, surpassing the performance of the present detector which reached a time resolution of 67~ps.

\subsection{Highly granular calorimeters}
\label{ssec:HighGranularity}

The availability of affordable SiPMs with good performance, combined with advances in microelectronics, has opened up new possibilities in calorimetry, making scintillator-based detector systems with millions of readout channels a credible option. This allows the construction of highly granular "imaging" calorimeters for collider experiments, which are optimized for event reconstruction with particle flow algorithms. Such detectors, primarily studied in the context of the ILD, SiD and CLIC detectors \cite{Behnke:2013lya, Linssen:2012hp} at the future linear colliders ILC and CLIC, would have on the order of 10 million SiPMs in the hadron calorimeter. The CALICE AHCAL physics prototype, discussed in \autoref{sec:Pioneers}, is the first scintillator / SiPM based imaging calorimeter developed in that context. 

This detector concept has been extended to an electromagnetic calorimeter with 10~mm wide, 45~mm long and 3~mm thick plastic scintillator strips with a wavelength-shifting fiber inserted in an extruded 1.5 mm diameter channel in the scintillator \cite{Francis:2013uua}. One end of the fiber is mirrored, while the other end is coupled via an air gap to a SiPM manufactured by Hamamatsu inserted in a notch at the end of the scintillator strip. The scintillator area around the fiber end in the notch is covered by a screening foil, which is crucial for achieving a uniform response along the strip by blocking photons from reaching the sensor without being first absorbed in the wavelength-shifting fiber, which would otherwise result in larger signals for particles passing the scintillator close to the SiPM. The SiPMs themselves are mounted on a custom-designed flex cable which connects the sensors to the front-end electronics. An effective granularity of $10 \times 10~\mathrm{mm}^2$ is achieved by crossing alternating detector layers. The scintillator elements are inserted in a tungsten carbide absorber structure with 3.5~mm thick absorber plates. A full prototype with 30 sampling layers, a thickness of 21.5 $X_0$ (266~mm), lateral dimensions of $180 \times180~\mathrm{mm}^2$ and a total of 2,160 scintillator strips and photon sensors was constructed and extensively tested in electron and hadron beams \cite{Repond:2017ccd}, also together with the CALICE AHCAL.

\begin{figure}
\centering
\includegraphics[width = 0.75\columnwidth]{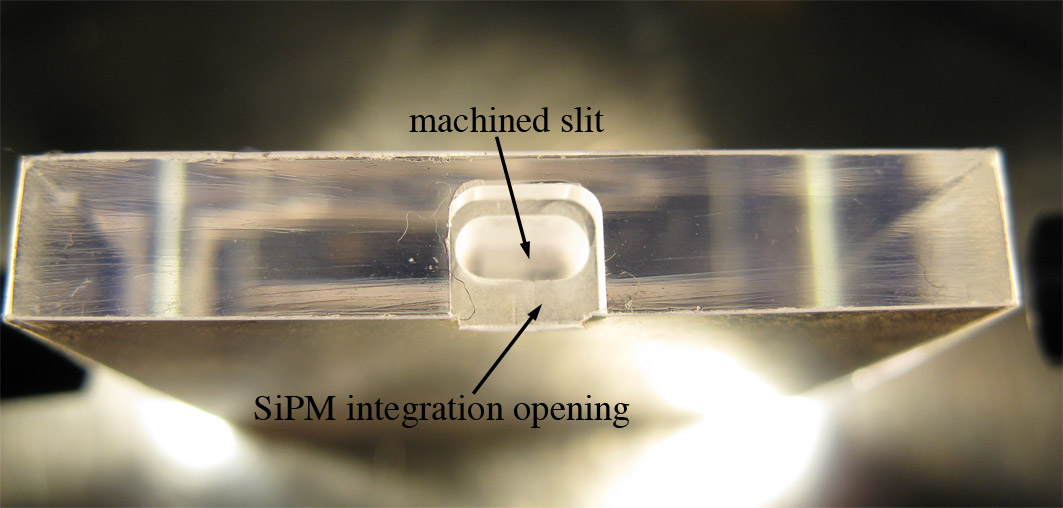}
\caption{A $30 \times 30 \times 5$~mm$^3$ plastic scintillator tile for the CALICE AHCAL optimized to achieve a high degree of response uniformity over the full active area with direct coupling of a SiPM to the scintillator. Figure taken from \cite{Simon:2010hf}\label{fig:Calorimetry:SideDimple}}
\end{figure}

The improvement of the performance characteristics of commercially available SiPMs has enabled a further evolution of scintillator-based highly granular calorimeters. Modern sensors have their maximum sensitivity in the blue spectral range around 420~nm, coinciding with the peak of the emission spectra of fast polystyrene and polyvinyltoluene-based plastic scintillators commonly used in high-energy physics. This eliminates the need for wavelength shifters, making a direct coupling of the SiPM to the scintillator possible. However, a wavelength-shifting fiber embedded in the scintillator also serves to collect the scintillation light, supporting a uniform response over the full area of the scintillator element. To preserve this important feature also in the absence of a fiber, specific geometries at the coupling position of the photon sensor, such as ``dimples'' \cite{Blazey:2009zz, Simon:2010hf} are needed.  Figure \ref{fig:Calorimetry:SideDimple} shows a scintillator tile optimized for a high degree of response uniformity when directly coupling a SiPM to one side of the scintillator. This design preserves the scintillator tile and the photon sensor as one integrated unit, allowing testing for quality assurance purposes prior to the installation of the photon sensor on the electronics board. The absence of a fiber also relaxes the positioning tolerances of the SiPM, which otherwise has to be carefully aligned with the fiber end to ensure a homogeneous number of SiPM pixels that are illuminated by the fiber for a large ensemble of tiles.  A key parameter of the tile / SiPM combination is the ``light yield'' for minimum-ionizing particles (MIPs), given by the most probable value of the number of detected photons. This value needs to be sufficiently high to ensure high efficiency and low sensitivity to the precise value of noise thresholds, and low enough to preserve an acceptable dynamic range. For the CALICE AHCAL, target values are on the order of 15 to 20 photons per MIP. A response on this level allows for a clean identification of MIPs for calibration purposes also within hadronic showers, where track segments of secondary hadrons can be reconstructed \cite{Adloff:2013vra}. The optimization of the number of photons detected per MIP is however application-specific. It depends strongly on the environment in which the detector will be operated in, which determines the expected range of signal amplitudes as well as the scintillator and photon sensor degradation during operation due to radiation damage, and on the efficiency and resolution requirements. 

\begin{figure}
\centering
\includegraphics[width = 0.75\columnwidth]{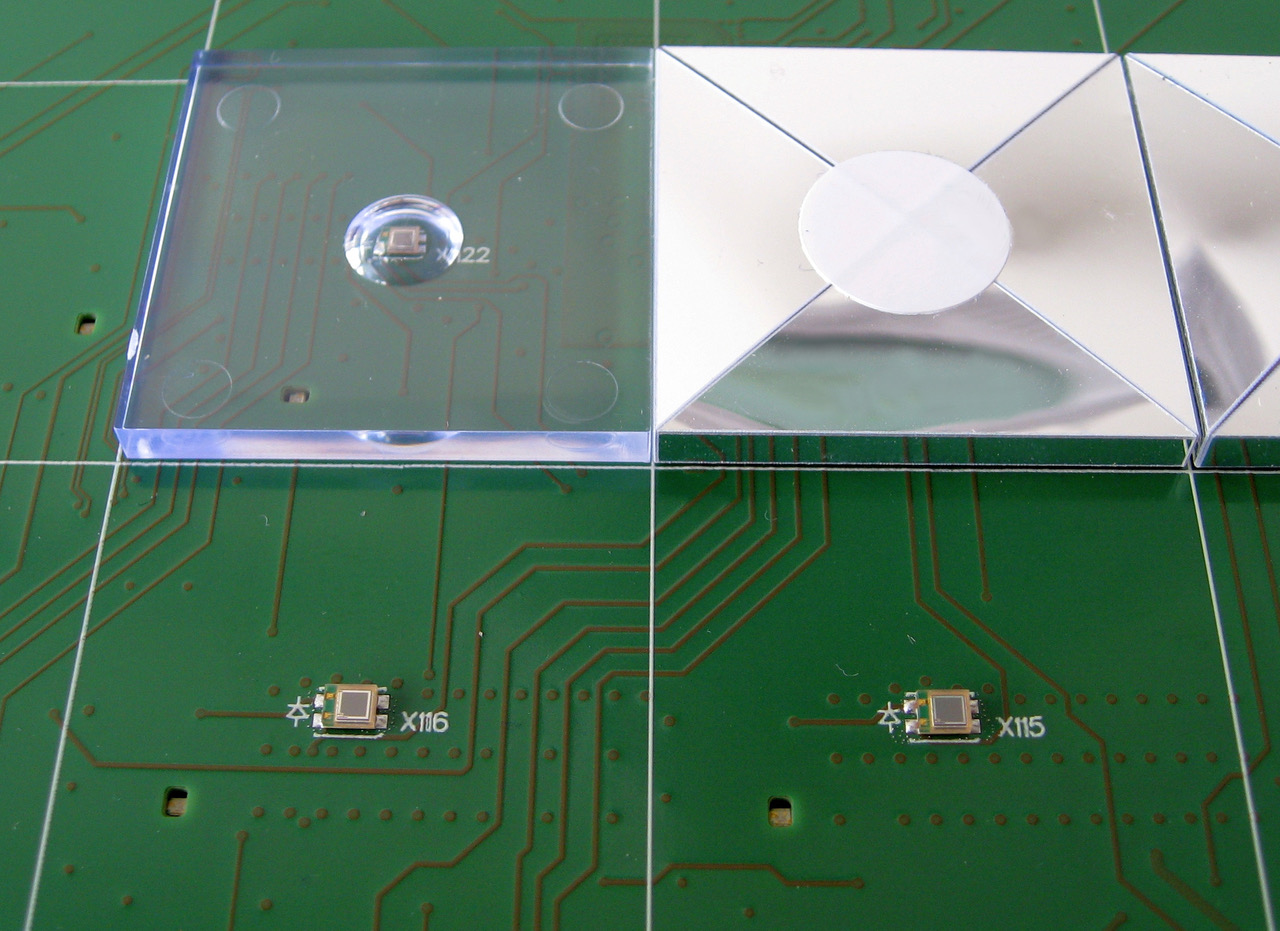}
\caption{Scintillator tiles with a dimple mounted on SiPMs on a PCB for the CALICE AHCAL technological prototype. The bare scintillator tile is shown for illustration, in the fully assembled detector all scintillator tiles are wrapped in reflective foil and glued onto the PCB. Figure taken from \cite{Sefkow:2018rhp}. \label{fig:Calorimetry:AHCAL_TPTile}}
\end{figure}

The advent of SiPMs in packages suitable for automatic surface mounting, combined with improvements in device-to-device uniformity and reliability has substantially simplified the construction of active elements for imaging calorimeters. The technical prototype of the CALICE AHCAL \cite{Sefkow:2018rhp} makes full use of these possibilities. With fully embedded electronics, SMD - style photon sensors and plastic scintillator tiles optimized for direct coupling \cite{Liu:2015cpe}, the active elements of the detector can be assembled almost fully automatically, and with a degree of response uniformity perfectly sufficient for hadron calorimetry. In contrast to the scintillator design shown in \autoref{fig:Calorimetry:SideDimple}, this design no longer allows the testing of individual SiPMs together with scintillator tiles prior to the full integration of the basic high-level detector elements. These elements have a size of $360 \times 360~\mathrm{mm}^2$, hosting 144 photon sensors controlled by 4 ASICs, and 144 30 $\times 30 \times 3~\mathrm{mm}^3$ scintillator tiles.  \autoref{fig:Calorimetry:AHCAL_TPTile} shows a close-up of this tile-on SiPM design. The high reliability and uniformity of the latest generation of SiPMs allows to do spot testing of bare SiPMs only prior to assembly, rather than a full characterization of each sensor together with the matching scintillator tile. The scintillator tiles use polystyrene-based scintillating material and are produced via injection moulding. The typical amplitude of 15 detected photons per MIP ensures high efficiency for single particles and a robust MIP-based calibration. With the electronics operated in a mode optimized for linear collider beam structures, the detector is capable of single cell time stamping on the nanosecond level for MIP-type signals. 

A full hadronic calorimeter prototype with a volume of approximately 1~m$^3$ and 22,000 SiPMs has been constructed based on this design \cite{Sefkow:2018rhp}, making use of automatized testing and assembly techniques. In May 2018, the detector was first exposed to high-energy particles at the CERN SPS, with just 8 dead channels (0.04\%), underlining the quality of present-day SiPMs as well as the robustness of the automated assembly techniques. 

\begin{figure}
\centering
\includegraphics[width = 0.95\columnwidth]{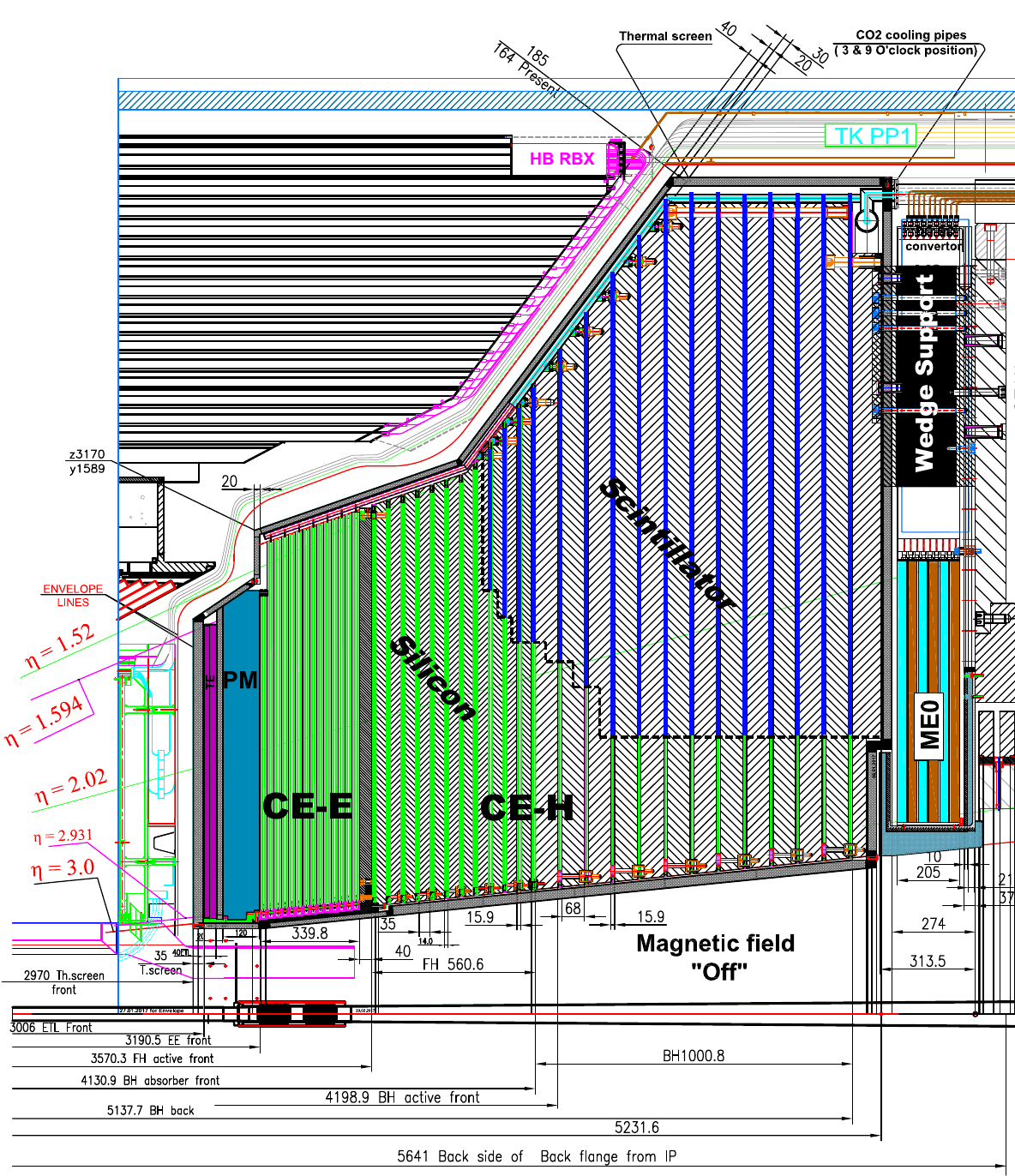}
\caption{Longitudinal cross section of the upper half of one CMS endcap calorimeter after the phase II upgrade, showing the electromagnetic (CE-E) and hadronic (CE-H) sections of the detector. The hadronic section uses both silicon and plastic scintillator with SiPMs as active elements. Figure taken from \cite{Collaboration:2293646}. \label{fig:Calorimetry:CMS_HGCAL}}
\end{figure}

This ``tile-on-SiPM'' technology has also been adopted for the scintillator part of the phase II upgrade of the CMS endcap calorimeter system \cite{Collaboration:2293646}. The technical design of the detector has recently been documented in its technical design report, with substantial R\&D still ongoing at present, and correspondingly the potential for larger design changes. Here the concept presented in the technical design report of the detector is discussed. This highly granular sampling calorimeter, illustrated in \autoref{fig:Calorimetry:CMS_HGCAL}, consists of an electromagnetic and a front and back hadronic section. It uses silicon pad sensors and scintillator tiles read out with SiPMs, the latter to reduce the overall cost of the system. Silicon is used throughout the electromagnetic section, in the first few layers of the front hadronic section as well as in the inner regions of the detector all the way to the last layer. In total, 600~m$^2$ of silicon sensors and 500~m$^2$ of scintillator elements with a total of 400,000 SiPMs will be used in the system. The transition between the the silicon and scintillator parts is given by the radiation damage and the corresponding loss of light yield of the scintillator and the high dark rates of the photon sensors at the end of life of the detector after an integrated luminosity of 3,000~fb$^{-1}$ at the HL-LHC. The requirements are driven by the need for a signal to noise ratio of at least 5 for minimum-ionizing particles to maintain the cell-to-cell calibration of the detector, which is needed to adjust the calibration as the radiation damage progresses.  Overall, this calorimeter upgrade will address the requirements imposed by the physics program at the HL-LHC, providing the capability for the separation of close-by particle showers and the minimization of the impact of pileup interactions on signal reconstruction thanks to its high lateral and longitudinal granularity, and ensuring adequate performance also at the end of the HL-LHC phase.

A key challenge of this detector is the severe radiation environment, which results in substantial radiation damage of the photon sensors. To keep the dark rate at a tolerable level, the sensors need to be cooled to $-30~^\circ$C. In addition, devices with a low pixel capacitance and thus a small pixel size and a fast recovery time on the order of 6 ns will be used to retain a sufficient photon detection efficiency, given by the number of pixels fully recovered and thus capable of being fired by an incoming photon at any given time, despite the high dark count rate up to the GHz level for a sensor with a size of 6~mm$^2$. This high dark rate also results in a substantial current-induced power dissipation of the photon sensors which reaches up to 20~mW per channel in the most irradiated regions. This makes the SiPMs themselves a key heat source in the scintillator layers, requiring a specific layout of the cooling system which provides cooling of the photon sensors through the PCB they are mounted on. 

Ongoing R\&D will establish the final design of the detector, including the precise scintillator geometries, the types of SiPMs to be used, and a decision on the possibility of coupling two sensors to the tiles in the most irradiated regions. Installation of the detector is foreseen in 2025 during Long Shutdown 3 of the LHC.

\subsection{Further calorimeter applications}

Beyond the applications discussed in some detail above, SiPMs are used in a variety of different calorimeter prototypes, and are considered for a number of future experiments and upgrades. It is impossible to provide an exhaustive list - instead a somewhat arbitrary collection of recent examples is briefly presented here. 

For the upgrade of the KLOE experiment for the KLOE-2 run beginning in 2014, two SiPM based calorimeters were constructed and installed. These are a 96 channel LYSO calorimeter, with each crystal read out by a large-area $4 \times 4~\mathrm{mm}^2$ SiPM \cite{Cordelli:2013mka}, and a 1,760 channel scintillator tile calorimeter with absorber plates made of a tungsten - copper alloy, based on 5~cm long scintillator tiles \cite{Balla:2013rka}. Each of the tiles has an embedded wavelength-shifting fiber inserted in a machined groove, read out by a SiPM.

SiPMs also enable substantial improvements in dual readout calorimetry based on scintillating and clear plastic fibers embedded in absorber structures. The possibility to locate the photon sensors directly at the position where the fibers leave the absorber structure eliminates the need to guide fiber bundles to PMTs. This allows a substantially more compact construction. In addition, it removes the issue that fiber bundles outside of the absorber structure may pick up particles not related to development of the shower inside of the calorimeter, resulting in an oversampling of late-developing showers. A first small prototype with a lateral size of $15 \times 15~\mathrm{mm}^2$  has demonstrated the potential of using SiPMs for such calorimeters, separately reading out the scintillation light produced in the scintillating fibers and the Cherenkov light produced in the clear plastic fibers with one SiPM coupled to each fiber end \cite{Antonello:2018sna}. In total 64 fibers, 32 of each type, were used in the prototype. The readout of each individual fiber brings high transverse granularity to fiber-based calorimeters. 

For the sPHENIX experiment at the Relativistic Hadron Collider, two SiPM-based calorimeters are currently being developed \cite{Aidala:2017rvg}, a SPACAL - style electromagnetic calorimeter with scintillating fibers embedded in a tungsten powder epoxy matrix, and a steel / scintillator-tile hadron calorimeter with wavelength-shifting fibers embedded in the scintillator plates. For the electromagnetic calorimeter, light guides are used to collect the light of a calorimeter tower on a $2 \times 2$ matrix of SiPMs. For the HCAL, the photon sensor is coupled to the wavelength-shifting fiber by means of a plastic structure which guarantees an air gap of 0.75~mm between fiber end and SiPM which has resulted in the best performance. For both calorimeters, SiPMs with $15\times 15$~$\mu$m pixels and an active area of $3 \times 3~\mathrm{mm}^2$ manufactured by Hamamatsu are used. The tests of a prototype of the full system show that the performance required for sPHENIX can be met with the current design.

\section{Applications in tracking detectors}
\label{sec:Tracking}

The small size and their insensitivity to magnetic fields make SiPMs also an attractive option for scintillator-based tracking systems. The properties of the SiPMs allow the placement of the photon sensors directly outside of the sensitive region of such a detector without the introduction of dead areas or the need for transportation of the scintillation light to regions with low magnetic field intensity. 

The potential of SiPM-based fiber trackers was first demonstrated in the context of balloon experiments \cite{Beischer:2010zz}, where also the relatively low operation voltage is a major advantage. Subsequently, this concept has been adopted for the upgrade of the LHCb experiment at the LHC at CERN \cite{LHCbCollaboration:2014tuj}, with a large scintillating fiber tracker with a total active area of 340 m$^2$ to be installed in the shutdown of 2019 and 2020.

\begin{figure}
\centering
\includegraphics[height = 0.9\columnwidth, angle=90]{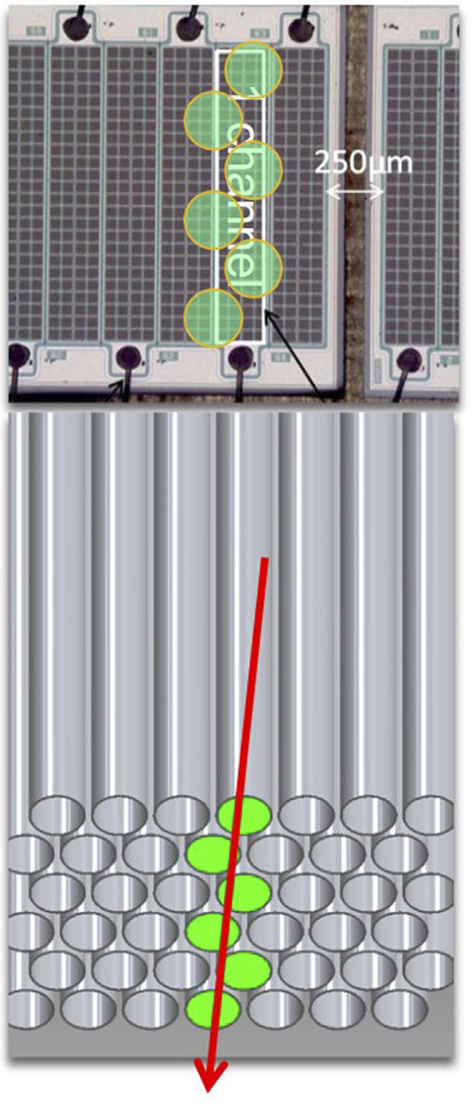}
\caption{Illustration of the working principle of the LHCb fiber tracker, with a photograph of a sub-region of the SiPM array shown on the left, together with an illustration of the fiber mat layout and the light collection pattern on the photon sensor. Figure taken from \cite{Kirn:2017kbn}. \label{fig:Tracking:LHCb}}
\end{figure}

The LHCb fiber tracker uses ``fiber mats'', which consist of 6 staggered layers of scintillating fibers, glued together by epoxy glue loaded with TiO$_2$ to reduce optical crosstalk between adjacent fibers. The fibers are 2.5~m long with a diameter of 250~$\mu$m, resulting in a thickness of the mats of 1.35~mm. One end of the fiber mat is coupled to Hamamatsu-made SiPM arrays, while the other end is mirrored to increase the light yield. The SiPM arrays are composed of two dies with 64 channels each. The pixel size of the sensors is $57\times 62~\mu$m$^2$, with 104 pixels per SiPM cell. Each SiPM cell is 250~$\mu$m wide, matching the diameter of the fibers and with that also approximately matching the fiber pitch, and 1.62~mm high, covering the full thickness of the fiber mat.  A sub-region of such an array, with several SiPM cells visible, is shown in \autoref{fig:Tracking:LHCb}. The full detector will have a sensitive area of 340~m$^2$, and will use more than 590,000 SiPMs.

The position of a throughgoing particle is reconstructed from the center of gravity of the observed light signals on several adjacent SiPMs, as illustrated in \autoref{fig:Tracking:LHCb}. Due to the staggering of the fibers, and since the fibers are not explicitly aligned with the SiPMs within the array, the light of one fiber is typically collected on two SiPMs, and multiple fiber contribute to the signal of a SiPM. A spatial resolution of $\sim$50~$\mu$m is achieved in test beams when using the full amplitude information. With the 2 bit readout used in LHCb, 80~$\mu$m are obtained \cite{Greim:2017hmn}.

The peak emission wavelength of the fibers of 450~nm fits well to the broad efficiency maximum of the photon sensors around 490~nm, where they reach a photon detection efficiency of 50\% with an overbias of 3.5~V.  The SiPMs have a low inter-pixel crosstalk of 5.5\% including both direct and delayed crosstalk. To keep afterpulsing to a negligible level a high quench resistor is used, resulting in a long recovery time, larger than integration time of the electronics. This is required to prevent a deterioration of the position resolution based on signal sharing by additional pulses not directly caused by scintillation light.

A key challenge is the radiation environment the sensors are operated in the LHCb experiment, with an expected ionisation dose of 80~Gy and a neutron fluence of $10^{11}$~n$_{\mathrm{eq}}$/cm$^2$. This will result in an increase of the dark count rate by approximately three orders of magnitude, up to tens of MHz per channel at room temperature. While this is substantially less severe than the radiation damage experienced by the SiPMs used in the CMS phase II endcap calorimeter discussed in \autoref{ssec:HighGranularity}, a much lower noise rate is required here in order to retain the capability for precise position resolution by light sharing over multiple SiPMs. By cooling the sensors to $-40$ $^\circ$C using a single-phase Novec 449 - based cooling system, the dark rates will be kept well below 1 MHz also at the end of LHC run 3 \cite{Kirn:2017kbn, Greim:2017hmn}, retaining the tracking precision of the system.

A substantially smaller system building on similar technology is being constructed for the Mu3e experiment \cite{Blondel:2013ia}, where a cylindrical three-layer scintillating fiber detector with a radius of 64 mm will be used for tracking and precision timing \cite{Bravar:2017ush}, complementing the highly precise but slower silicon pixel tracking layers. The light readout will be based on the LHCb-type multi-channel SiPM arrays, with a readout on both detector sides to achieve a time resolution below 500~ps. In total, the detector has 3,072 channels, split over 24 arrays with 128 SiPMs each. A system with two crossed 32~mm wide scintillating fiber ribbons, each read out by one LHCb-type SiPM array, has been constructed for the NA61/SHINE beam position monitoring system \cite{Damyanova:2017gwv}, to be installed in 2019.

The coupling of scintillating fiber mats to SiPM arrays as done for the examples discussed above offers the advantage of an easy and compact construction with moderate requirements on alignment tolerances. It also offers high particle detection efficiencies since typically the light of several fibers is collected on one SiPM, resulting in larger signals and thus a reduced impact of noise thresholds. Studies have shown that slightly higher spatial resolutions can be reached with a fiber-by-fiber readout, at the expense of a lower efficiency and substantially increased mechanical complexity \cite{Damyanova:2017gwv}.

Scintillator-based tracking detectors are not limited to fiber detectors. They also make use of scintillator bars, for example for muon identification and tracking  systems discussed in \autoref{sec:ParticleID}, or of other scintillator elements. One particularly interesting example is the upgrade of the T2K ND280 \cite{Blondel:2299599}. Here, a three-dimensional scintillator tracker, the so-called SuperFGD, is in development. This system is based on densely packed 1~cm$^3$ scintillator cubes with three orthogonal wavelength-shifting fibers inserted in aligned channels in the scintillator matrix, read out on both ends by SiPMs \cite{Sgalaberna:2017khy, Mineev:2018ekk}. The detector serves as an active target for the neutrinos, and provides tracking capabilities over the full solid angle. A 2.2~m$^3$ system will use 58,800 photon sensors. The same technology is also discussed as a possible component of the near detector of the DUNE experiment.

\section{Applications in particle ID and veto systems}
\label{sec:ParticleID}

The application of SiPMs in particle identification and veto systems can be roughly grouped into three classes of detector technologies: Cherenkov detectors where SiPMs are employed to directly detect the Cherenkov light, plastic scintillator-based charged particle and shower detectors where SiPMs are used to detect scintillation light, and time-of-flight detectors. The latter application is discussed together with other SiPM-based timing systems in \autoref{sec:Timing}. Particle ID and veto systems based on plastic scintillators are technologically very closely related to the SiPM applications in calorimetry discussed in \autoref{sec:Calorimetry}, while Cherenkov detectors make use of the capability for blue and UV sensitivity of recent generations of SiPMs, with requirements similar to those imposed by many astro-particle physics experiments, as discussed in detail in \cite{RazmikSiPM}.

\subsection{Detection of Cherenkov light}

Even though SiPMs are in principle well suited also for the detection of Cherenkov photons, they are to date not yet used in ring-imaging Cherenkov detectors in larger running particle physics experiments. Such detectors require the coverage of a  large area with photon detectors. A key challenge when using SiPMs is the relatively high dark count rate of the sensors. Since ring imaging Cherenkov detectors rely on the detection of rings with a few photons, single photon detection with high efficiency and purity is required. To achieve this with SiPMs, the needed sensor area is minimized by using light concentrators in front of the photon sensors.  This increases the geometrical acceptance and thus the signal level, while keeping the silicon surface small, which reduces the total dark rate. In addition, narrow time windows on the level of a few ns can effectively reduce accidental coincidences of true photons and dark pulses. 

With these techniques, promising results have been achieved in the context of the developments for the Belle II aerogel RICH detector \cite{Korpar:2010zza}, demonstrating that SiPMs could in principle replace other types of photon detectors used in current RICH detectors, such as the hybrid avalanche photo detectors (HAPDs) that were selected for the Belle II ARICH \cite{Pestotnik:2017hrk}. While HAPDs are capable of working in perpendicular magnetic fields, as found in the endcap region of collider experiments, SiPMs provide increased flexibility by allowing an arbitrary field strength and orientation. However, the limited radiation tolerance of SiPMs presents a major challenge for such devices, since the detection of single photons is quickly deteriorated by increasing dark rates due to bulk damage of the sensors. This made the use of present-generation SiPMs in the Belle\,II  ARICH impossible.

\subsection{Scintillator readout}

The Belle II experiment does use SiPMs in the endcap $K_L$ and muon detector (EKLM) \cite{Aushev:2014spa}, a system that merges particle identification capabilities and calorimeter features. In this system, extruded scintillator bars arranged as crossed double layers are used as detection medium, sandwiched by 4.7~cm thick iron plates in the flux return of the Belle II solenoid. The scintillation light is collected by a wavelength-shifting fiber glued into a central groove machined into the 7~mm thick, 40~mm wide and up to 3~m long polystyrene-based scintillator strip. This groove is rounded at the bottom to minimize residual air bubbles, resulting in an increased light yield compared to a standard rectangular groove. One end of each fiber  is coupled to a Hamamatsu-made SiPM with an active area of $1.3 \times 1.3~\mathrm{mm}^2$ and a pixel pitch of 50~$\mu$m, while the other end is mirrored \cite{Uglov:2017gsz}. The readout on only one side of the scintillator strip, dictated by the detector geometry, results in a strong dependence of the light yield on the distance of the particle impact from the photon sensor, with a factor of two reduction in signal observed for particles at a distance of 3~m compared to an incidence close to the light sensor \cite{Aushev:2014spa}. This can be corrected for in the data analysis, making use of the crossed strip geometry in alternating layers of the detector.

\begin{figure}
\centering
\includegraphics[width = 0.99\columnwidth]{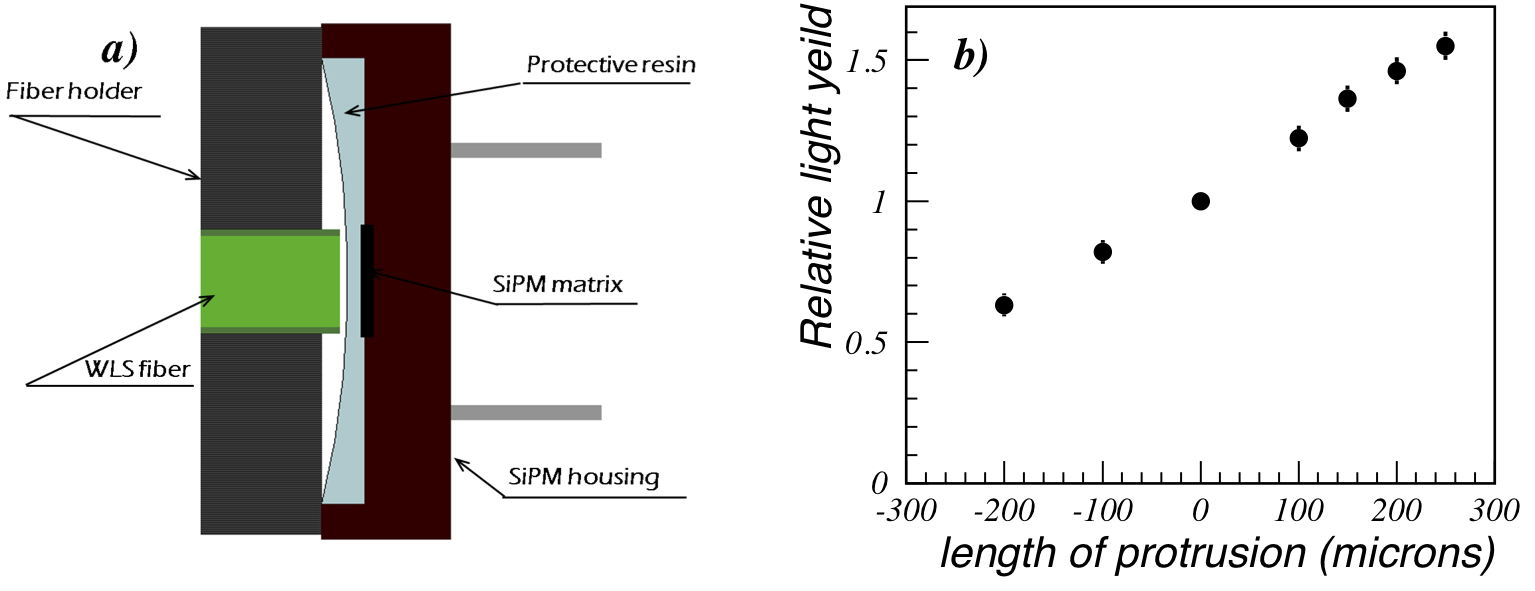}
\caption{a) Schematic view of the coupling of the wavelength-shifting fiber embedded in the Belle II KLM scintillator bars to the photon sensor showing the concave shape protective resin and b) the relative light yield as a function of the protrusion of the fiber end below plane given by the outer rim of the photon sensor package. Figure taken from \cite{Aushev:2014spa}. \label{fig:ParticleID:KLM}}
\end{figure}

 A key factor in the optimization of the signal yield of the detector is the optical coupling of the wavelength-shifting fiber to the SiPM. The sensor is covered by a protective resin, which forms a concave surface due to the effect of surface tension during the hardening of the resin. This shape results in a defocusing of the light from the fiber on the SiPM. Optimal coupling and light yield is reached when minimizing the distance of the fiber end to the resin surface in the center of the sensor, as illustrated in \autoref{fig:ParticleID:KLM}.
In total, 15,600 strips are installed in the EKLM system, which started operation with beam in the first collision run of Belle II in 2018. A muon system based on the same technology is also considered for the SHiP experiment \cite{Anelli:2015pba}.

\begin{figure}
\centering
\includegraphics[width = 0.99\columnwidth]{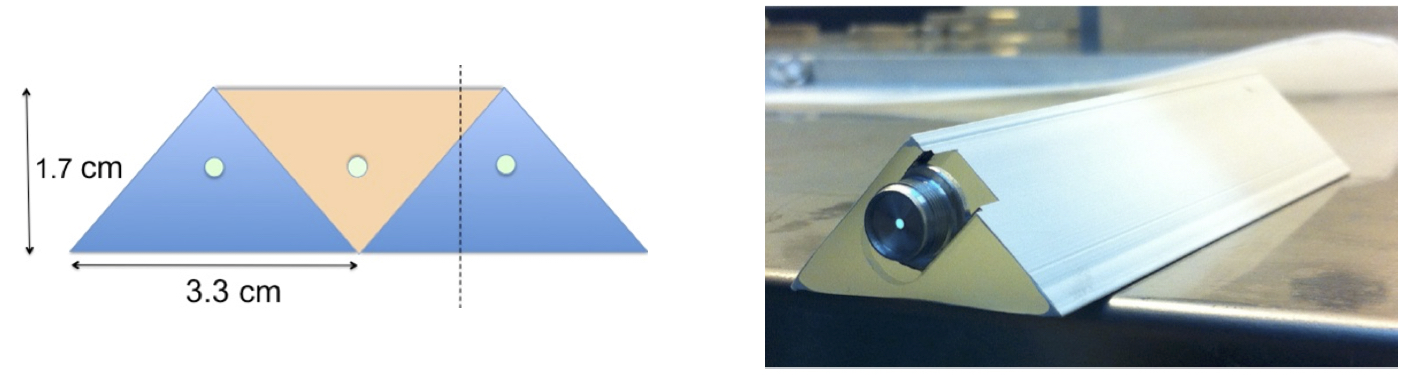}
\caption{Left: Arrangement of triangular scintillator bars with embedded wavelength-shifting fiber for the CHANTI detector of NA62, also illustrating the position-dependent light sharing between two neighboring strips for a penetrating particle indicated by the vertical dashed line. Right: One triangular scintillator bar, showing the coupling connector for the photon sensor. Figure taken from \cite{NA62:2017rwk}. \label{fig:ParticleID:NA62}}
\end{figure}

The charged anti-coincidence detector (CHANTI) of the NA62 experiment \cite{NA62:2017rwk,Ambrosino:2015ruf}, used to reject background particles from beam interactions in detector elements, uses  triangular polystyrene-based scintillator bars extruded with a central hole for the insertion of a wavelength-shifting fiber for light collection and a co-extruded TiO$_2$ coating. The bars, with a maximum length of 30~cm, are read out on one end, with the other end of the fiber mirrored. The triangular bars with a base width of 33~mm are arranged as shown in \autoref{fig:ParticleID:NA62}, resulting in a plane detector with a thickness of 17~mm. This arrangement results in sharing of the signal of through-going particles over two neighboring strips, which is used to improve the spatial resolution to 2.5~mm by making use of the energy sharing between strips. The same type of SiPM as for the Belle II EKLM discussed above is used for the readout of the fibers. A dedicated connector consisting of precision-machined aluminum parts guarantees good and stable optical coupling between fiber and photon sensor. The body of the connector is glued to the scintillator and fiber, while a screw cap holds the SiPM in place \cite{Ambrosino:2015ruf}. The total system uses 288 scintillator strips.

\section{Applications in large-volume detectors}
\label{sec:LargeVolume}

Large volume detectors, used for example in long baseline neutrino oscillation experiments, have quite different requirements than most of the other applications discussed in this review. The large volumes imply large areas to be covered by the photon detectors, often with very strict limits on the noise level. The planned next generation of large water Cherenkov detectors such as HyperK \cite{Abe:2018uyc} and the liquid scintillator - based detector of the JUNO experiment \cite{Djurcic:2015vqa} currently under construction thus both use photomultiplier tubes for light readout. 

The single-phase liquid argon TPCs in the DUNE experiment \cite{Acciarri:2016ooe} will use SiPMs in the photon detection system. In DUNE, the primary purpose of the photon detection system is to provide a trigger and a precise starting time for neutrino interactions in the liquid argon, which are of particular importance for the reconstruction of non-beam events. Since the event reconstruction including the determination of the energy is primarily based on the reconstructed ionisation in the liquid argon, the requirements on the overall coverage and photon collection efficiency of the system are substantially reduced compared to water and liquid scintillator-based detectors. The cryogenic operation results in very low dark count rates, largely eliminating noise concerns. The low temperatures require the use of sensors with a metal quenching resistor, since the large temperature dependence of polysilicon resistors results in excessively long tails in the signal pulses \cite{Otono:2006zz, JanicskoCsathy:2010bh}.

\begin{figure}
\centering
\includegraphics[width = 0.6\columnwidth]{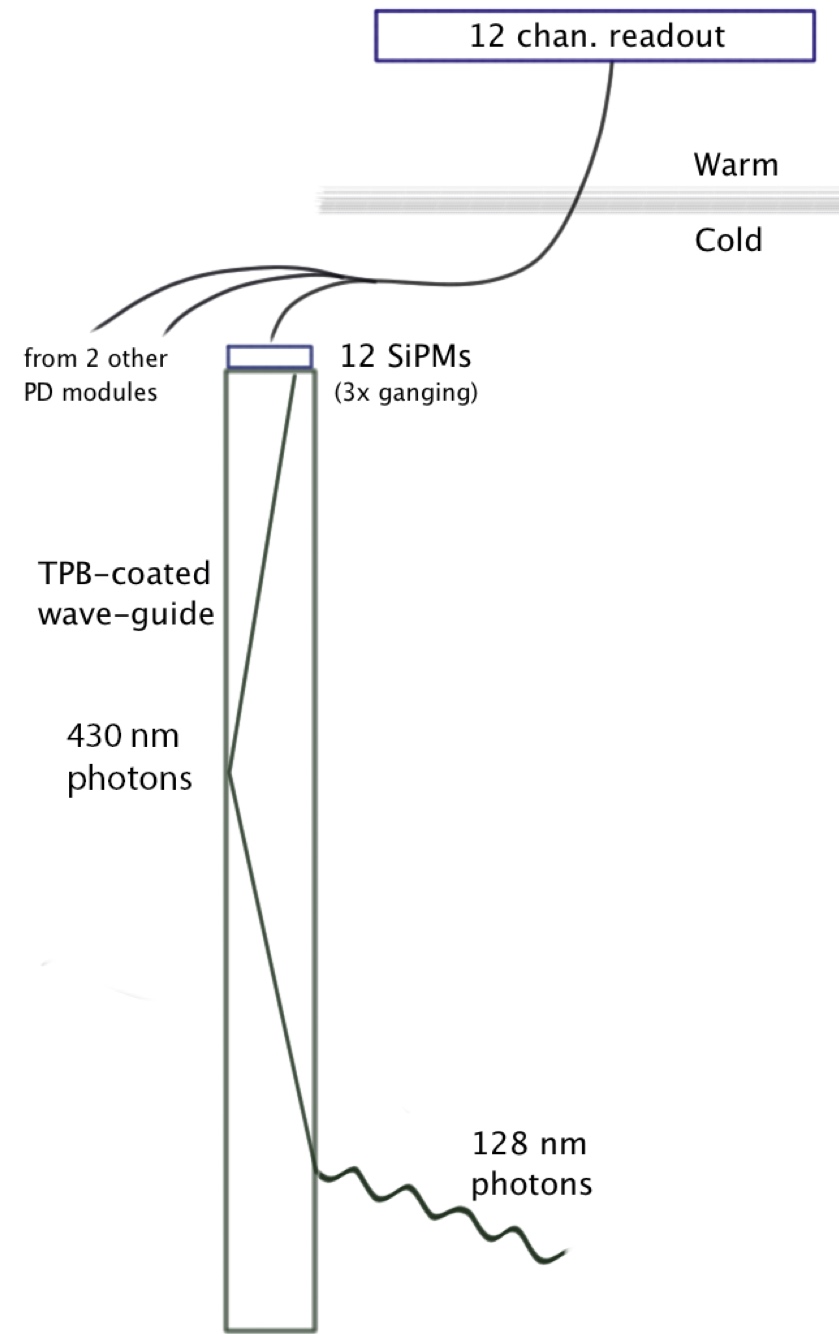}
\caption{Illustration of the reference design of the photon detectors for the DUNE single phase liquid argon TPCs. The TPB-coated light guide and the SiPMs are located inside the liquid argon, with feedthroughs needed in the cryostat to guide the analog signals to the electronics outside. Figure taken from \cite{Acciarri:2016ooe}. \label{fig:LargeVolume:DUNEPD}}
\end{figure}

Several different concepts are currently being developed for photon detection in the DUNE liquid argon TPCs. They all include wavelength-shifting to match the 128~nm VUV scintillation light from the liquid argon to the spectral sensitivity of the photon sensors, and provide photon collection over areas much larger than the active sensor area. Thus these concepts do not make use of special VUV sensitive SiPMs as discussed in the context of the MEG II liquid Xenon calorimeter in \autoref{ssec:ScintCryo}, but do require devices optimized for cryogenic operation. The reference design \cite{Acciarri:2016ooe} is based on a $2.2~\mathrm{m} \times 83~\mathrm{mm} \times 6~\mathrm{mm}$ acrylic light guide coated in TPB. The TPB serves as a wavelength-shifter for the VUV scintillation light, shifting it into the sensitive region of standard SiPMs. The bars are read out on one end by 12 SiPMs manufactured by SensL, each with an active area of $6 \times 6~\mathrm{mm}^2$.  The SiPMs are ganged in groups of three to limit the number of electronics channels in order to reduce the required feedthrough cross section for the cryostat. \autoref{fig:LargeVolume:DUNEPD} illustrates the design of one photon detector modules. These modules will be inserted in the Anode Plane Assemblies (APAs) that use wires to collect the drifting electrons from the liquid argon. This imposes strict constraints on the thickness of the photon detection system. Overall, the coverage of the photon detection system is rather sparse, with 10 bars inserted in a 6.3~m long unit. For one DUNE single phase liquid argon TPC this results in 1,500 photon detectors, with a total of 18,000 SiPMs. 

The photon detector concept was demonstrated in a 35 ton liquid argon TPC prototype, using different light guide and coating concepts, smaller than the final design and read out by 3 SensL SiPMs each \cite{Adams:2018lfb}. Larger-scale tests are imminent in the ProtoDUNE-SP detector, where 60 full-size photon detector modules are installed \cite{Abi:2017aow}. In addition to the reference design and variations thereof using different coating techniques or doping of the light guide, other options that may provide a higher photon capture efficiency are also being explored. These are based on photon traps, where the scintillation light is captured, wavelength-shifted and trapped in the detector volume, and then read out by SiPMs typically after multiple reflections. One of those is the ARAPUCA device \cite{Machado:2016jqe}, which uses dichroic filters to shift and trap photons in a highly reflective volume, where they are detected by SiPMs which only cover a very small area of the overall surface, on the level of 0.15\% for an expected photon detection efficiency of 1\%. First tests in liquid argon have shown a performance close to expectations \cite{Cancelo:2018dnf}, with more results expected from ARAPUCA modules installed in the ProtoDUNE detector \cite{Abi:2017aow}. A variation of the ARAPUCA concept is the ArCLight detector, which uses a wavelength-shifting plastic body instead of an unfilled reflective volume, resulting in higher mechanical stability and a more compact construction \cite{Auger:2017flc}. Here, the light is collected by SiPMs on the side of the light collector.  This device is currently being developed for use in ArgonCube liquid argon TPCs \cite{Auger:2268439} under consideration as one component of the DUNE near detector. A prototype ArCLight detector with a size of $10 \times 10~\mathrm{cm}^2$ is read out by for Hamamatsu-made SiPMs with an active area of $3 \times 3~\mathrm{mm}^2$ each has shown promising performance in first tests conducted at room temperature.

\section{Developments towards precision timing systems}
\label{sec:Timing}

The fast rise time of SiPM signals makes them well suited for timing detectors when coupled with fast scintillators or when detecting Cherenkov light. With suitable electronics, the sensors offer the potential for a time resolution of a few 10~ps for relatively modest signal amplitudes \cite{Gundacker:2013ywa}. At present, larger systems are not yet in use in experiments, but two classes of applications for timing of single charged particles are currently in active development. These are plastic scintillator-based timing, veto and time-of-flight systems, and systems based on inorganic scintillators. Examples for the former application are the timing detectors for the MEG II, Mu3e and SHIP experiments, the TOF system for PANDA, as well as detectors used to study the time structure of hadronic showers and to measure injection background in SuperKEKB. The largest project in the latter group currently in development is the proposed Barrel Timing Layer of the CMS experiment. In addition to these dedicated timing detectors, which primarily target minimum-ionizing particles, the capability for precision timing is also used in  other SiPM-based systems, such as calorimeters and tracking detectors, discussed in \autoref{sec:Calorimetry} and \autoref{sec:Tracking}, respectively.

\subsection{Plastic scintillator - based timing systems}
\label{ssec:Timing:Plastic}

The timing detector is the second SiPM-based system in MEG II. The small size and immunity to magnetic fields of SiPMs enable the upgrade of the scintillator bar / PMT based timing detector of the MEG experiment to the ``pixelated timing counter'' based on scintillator tiles with SiPM readout \cite{Baldini:2018nnn}. This new system can sustain the higher hit rate expected in MEG II thanks to a substantial increase in granularity, and is also expected to reach a better time resolution. 

\begin{figure}
\centering
\includegraphics[width = 0.65\columnwidth]{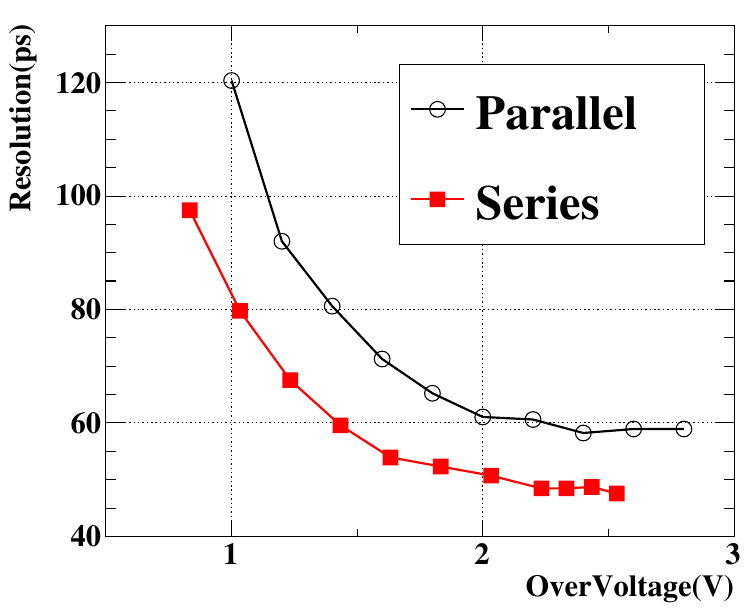}
\caption{Time resolution of a $60 \times 30 \times 5$ mm$^3$ scintillator tile read out with three Hamamatsu SiPMs connected either in series or in parallel as a function of overvoltage. Figure taken from \cite{Nishimura:2015qev}. \label{fig:Timing:MEG}}
\end{figure}

\begin{figure}
\centering
\includegraphics[width = 0.65\columnwidth]{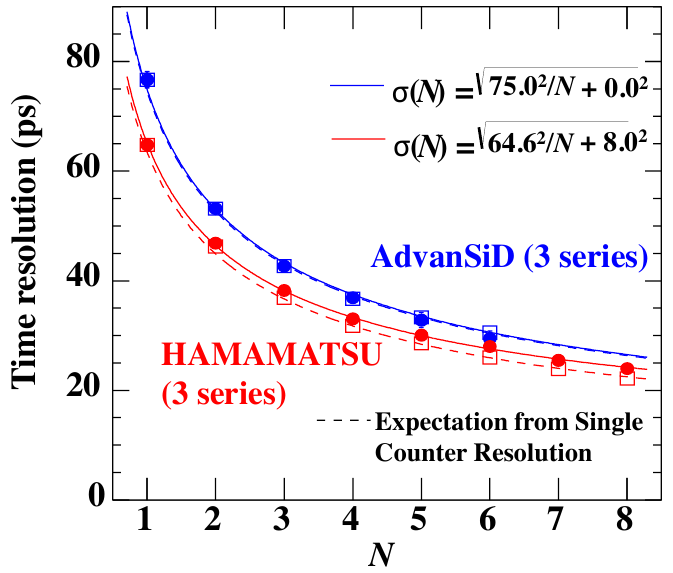}
\caption{Time resolution of  test beam systems of the MEG II timing detector as a function of the number of crossed detector elements, comparing Hamamatsu and AdvanSiD devices. Note that the scintillator tiles used in this test are smaller than the ones in the final experiment, resulting in better time resolution. Figure taken from \cite{Nishimura:2015qev}. \label{fig:Timing:MEGCompare}}
\end{figure}

The plastic scintillator tiles of the timing counter are 120~mm long, 40 or 50~mm wide and 5~mm thick. For an optimal time resolution, light readout on both short edges, with as much active coverage as possible, is required. Studies with different schemes of combining the signals of three SiPMs used on one side of the scintillator tile have shown that a connection in series provides substantially better time resolution than a connection in parallel \cite{Nishimura:2015qev}, as illustrated in \autoref{fig:Timing:MEG}. The connection in series results in a smaller capacitance, and thus faster rise times and shorter pulses overall. However, this scheme requires higher voltages, since the operation voltages of multiple sensors is are added together. 

For the choice of the SiPMs for the MEG II timing detector, devices from different vendors were considered. \autoref{fig:Timing:MEGCompare} shows the time resolution obtained in beam tests for prototype detectors as a function of the number of crossed $90 \times 40 \times 5$ mm$^3$ scintillator tiles, comparing the performance obtained with Hamamatsu and AdvanSiD SiPMs. The scintillator tiles are read out with three SiPMs per side, each with an active area of $3 \times 3~\mathrm{mm}^2$ and with 3,600 pixels. The time resolution is further improved with an increased number of SiPMs per scintillator tile. While the Hamamatsu devices provide a better performance for an equal number of sensors, the best overall time resolution within the given budget constraints was achieved with the AdvanSiD SiPMs, using 6 sensors per side of the tile coupled to the plastic by optical cement. These sensors are of ``NUV'' type with near-UV sensitivity well-matched to the emission spectrum of fas plastic scintillators. The adopted devices have a low breakdown voltage of around 24~V, which is an important advantage for passive series connection. They also show a low temperature dependence of the breakdown voltage \cite{Rossella:2017zwj}. A  total of 6,144 sensors will be installed in the detector. A first pilot run has demonstrated a single-tile time resolution of 93~ps, which improves to 31~ps for a mean number of hits of 9 \cite{Baldini:2018nnn} expected for a signal track.

\begin{figure}
\centering
\includegraphics[width = 0.8\columnwidth]{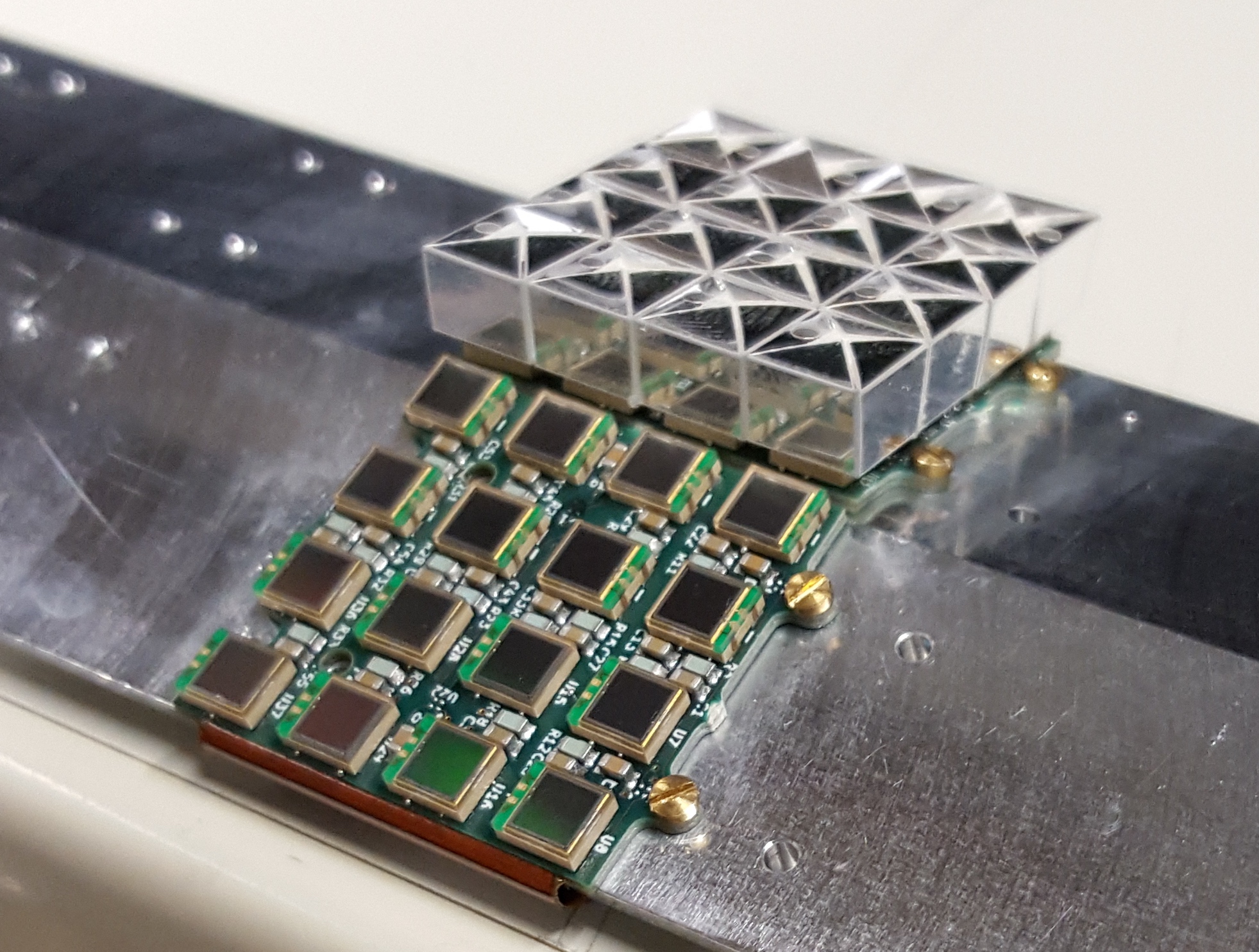}
\caption{A partially assembled sub-module of the Mu3e tile detector, showing the plastic scintillator tiles wrapped in reflective foil, and the arrangement of the $3 \times 3~\mathrm{mm}^2$ SiPMs used to read out the tiles. Figure taken from \cite{Munwes:2018}. \label{fig:Timing:Mu3e}}
\end{figure}

The Mu3e experiment \cite{Blondel:2013ia} uses two plastic scintillator-based timing detectors, the scintillating fiber tracker already discussed in \autoref{sec:Tracking}, and a scintillator tile detector \cite{Eckert:PhDThesis}. The tile detector provides precise time stamping of charged particles to detect coincidences of three electron / positron candidates, with the granularity of the detector allowing the unambiguous association to tracks measured in the tracker. To achieve this high granularity, the Mu3e timing detector uses a different layout than the detector of MEG II discussed above, with much smaller scintillator elements read out by a single photon sensor each. The detector consists of PVT-based plastic scintillator tiles with a size of $6.5 \times 6.0 \times 5.0~\mathrm{mm}^2$, wrapped in highly reflective foil. The scintillator type is Bicron BC-418, which provides a signal rise time of 0.5~ns and a pulse width of 1.2~ns. Each tile is directly read out by a $3 \times 3~\mathrm{mm}^2$ SiPM, thus a single sensor covers a considerable fraction of the total area of one edge of the scintillator tile. The detector consists of two barrels (``recurl stations'') on each side of the target, with a total of 6,272 channels, covering an area of 0.25~m$^2$. For a second phase of the experiment, it is planned to extend the detector by another two recurl stations. \autoref{fig:Timing:Mu3e} shows one partially assembled 32 channel sub-module of the detector, with wrapped scintillator tiles and the photon sensors visible. A time resolution of better than 100~ps is expected for the final system. In test beam measurements of a 16 channel prototype a time resolution of 56~ps has been achieved \cite{Eckert:PhDThesis}.   

Similar timing requirements also have to be met by the timing detector of the SHiP experiment \cite{Anelli:2015pba}. For this system with an area of 72~m$^2$, a total of 240 plastic scintillator bars with dimensions of $305 \times 11 \times 2~\mathrm{cm}^3$, read out by SiPMs are one of the options. First prototype studies with shorter scintillator bars and a single $3 \times 3~\mathrm{mm}^2$ SiPM on each end have shown a time resolution of 1~ns, indicating that a time resolution of 100~ps or better is achievable with larger photon-sensor coverage \cite{Betancourt:2017kkz}. 

The good time resolution of plastic scintillator / SiPM systems also lends itself to the application in time-of-flight detectors for particle identification. For the PANDA experiment such a system, which will cover an area of approximately 5.2~m$^2$, is in development for the barrel time-of-flight detector \cite{Gruber:2015moa}. Different geometries for the scintillator elements have been studied. The original proposal of tiles with a size of $30 \times 30 \times 5~\mathrm{cm}^3$, read out by two SiPMs on opposing edges has shown a relatively large position dependence of the time measurement, while a complete coverage of two opposing edges for the tiles, or the use of elongated rods or tiles, where the short edges are fully covered by photon sensors, result in a sufficiently homogeneous response \cite{Bohm:2016jhc}. The adopted design is a scintillator tile with a size of $87 \times 29.4 \times 5~\mathrm{mm}^3$, read out on both short edges by four $3 \times 3~\mathrm{mm}^2$ SiPMs each, similar to the solution adopted for the MEG\,II timing detector. The four SiPMs on each side are connected into a single channel. This design results in a total of 1,920 tiles with 13,460 SiPMs \cite{Suzuki:2018rxy}. Beam tests have demonstrated that the position dependence of the time response is negligible, and a time resolution below 60~ps can be reached.

\begin{figure}
\centering
\includegraphics[width = 0.9\columnwidth]{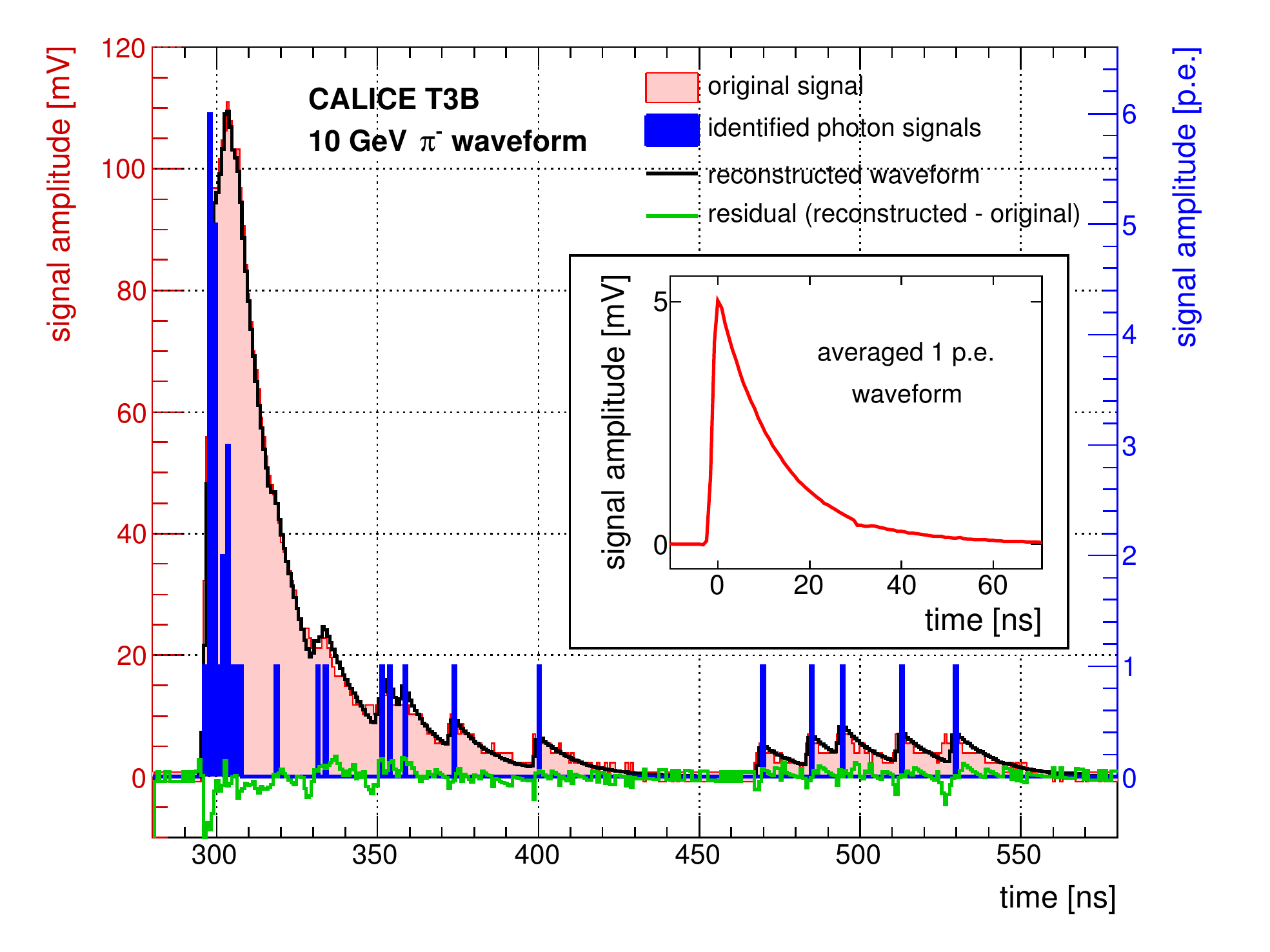}
\caption{Illustration of the waveform decomposition algorithm used in CALICE T3B to provide sub-ns time resolution on particle signals from an analysis of the analog waveform recorded from the SiPMs. The averaged waveform of a 1\,p.e.\
  signal (plot inset) is subtracted iteratively from the analog physics waveform (red) resulting in a $1\ \mathrm{p.e.}$ hit histogram
  (blue) representing the time of arrival of each photon on the light sensor. From this histogram, the original waveform is reconstructed back for cross
  checking purposes (black). The residual difference between the reconstructed and the original waveform (green) demonstrates the accuracy of the algorithm. Figure taken from \cite{Simon:2013zya}, copyright SISSA Medialab Srl. Reproduced by permission of IOP Publishing. All rights reserved. \label{fig:Timing:T3B}}
\end{figure}

A somewhat more exotic early application of the timing capabilities of SiPM - based systems was the study of the time structure of hadronic showers as part of the CALICE program discussed in \autoref{ssec:HighGranularity}, where a system with 15 scintillator tiles with embedded SiPMs read out with digitizers with deep buffers providing sub-ns time resolution for minimum-ionizing particles over a continuous readout period of more than 2 $\mu$s \cite{Simon:2013zya}. In this detector, referred to as CALICE T3B, the time of particle signals is reconstructed from the fully recorded analog waveform by a decomposition technique using an averaged single photon pulse, as illustrated in \autoref{fig:Timing:T3B}. The constant update of the single photon reference during running ensures an automatic correction for gain changes induced by temperature variations. 

From this system, a detector using to the latest generation of SiPMs and more capable digitizers was derived for the study of beam injection background during the commissioning of SuperKEKB. Here, minimum-ionizing particles and synchrotron radiation photons are measured with a continuous readout window of up to 100 ms, corresponding to 10000 revolutions of the accelerator, by the CLAWS detector consisting of 16 scintillator tiles with SiPM readout\cite{Gabriel:2016zlw, Lewis:2018ayu}. The sub-ns time resolution allows the association of the observed particles to a specific bunch in the SuperKEKB machine.

\subsection{Crystal - based timing systems}

For the Phase II upgrades of the LHC experiments, a new generation of timing detectors is in development. These detectors target a time resolution of 30 ps for minimum-ionizing particles to substantially reduce the impact of pileup. While the radiation levels in the endcap regions of ATLAS and CMS make silicon detectors such as LGAD sensors the only viable solution, fast scintillating crystals with SiPM readout, cheaper than all-silicon solutions, are possible for the CMS barrel system. 

The barrel of the CMS MIP timing detector \cite{Collaboration:2296612}, which is currently in its design phase towards a Technical Design Report, will cover an area of approximately 36.5 m$^2$ using LYSO crystals with SiPM readout. The detector has to fit within a 25~mm thick envelope between the CMS tracker and the electromagnetic calorimeter. It will consist of 250,000 LYSO tiles with a size of $11.5 \times 11.5$~mm$^2$ and a thickness of 3.7~mm to 2.4~mm depending on pseudorapidity. Each scintillator tiles will be read out by a SiPM with an active area of $4 \times 4$~mm$^2$. To increase the light yield and improve the time resolution, the scintillator is wrapped in reflective foil. The high radiation levels and corresponding expected damage of the SiPMs requires operation at $-30~^\circ$C using CO$_2$ cooling, as for the CMS endcap calorimeter upgrade discussed in \autoref{ssec:HighGranularity}. Still, the increase in dark count rate, and the corresponding need to increase thresholds, will result in a mild deterioration of the time resolution towards the end of HL-LHC operations. It is expected that the time resolution increases by approximately 60\% after a fluence of $2 \times 10^{14}\ \mathrm{n}_\mathrm{eq}/\mathrm{cm}^2$ \cite{Collaboration:2296612}.  

Test beam measurements have shown that sub-30 ps time resolution can be achieved with single LYSO crystals with SiPM readout \cite{Collaboration:2296612}. Achieving this time resolution is only possible with a correction for the position dependence of the response time of the crystal. This correction requires a precision of the particle impact point of better than 1~mm to keep the contribution of this effect to the time resolution below 20~ps. 

\section{General trends and optimization choices}
\label{sec:Trends}

The already large and diverse use of SiPMs in accelerator-based experiments in high energy and nuclear physics discussed here allows to draw some first conclusions concerning trends and directions in the use of these still rather new devices. 

The motivations for using SiPMs can be very coarsely divided into two main drivers: Detector technologies that are impossible or at the very least very impractical with other means of photon detection, and applications where SiPMs are used instead of other detectors because they provide better performance or other favorable features, cost savings or are considered the ``more modern'' solution. 

Not surprisingly, the first uses of SiPMs fall into the first category, since early generations of SiPMs still were rather expensive, had high dark count rates and other unwanted features, and did not yet achieve the levels of photon detection efficiencies reaching or surpassing those of conventional photomultipliers. This made their use attractive primarily for systems that would otherwise not have been feasible. Highly granular calorimeters on scintillator basis as developed by CALICE would not have been possible with optical fibers routing the scintillation light to photon detectors in regions well-shielded from magnetic fields based on space constraints in collider detectors, and for the fine-grained detectors in the T2K ND280 similar arguments apply. Here, SiPMs served as an enabling technology, offsetting the limitations of the early generations of sensors.

The continuous improvement of SiPMs, combined with a downward trend in prices per unit area, have opened up additional possibilities for application. The substantial reduction of unwanted effects such as afterpulsing and in particular inter-pixel crosstalk, the latter achieved by the introduction of optical barriers (``tenches'') between microcells, suppress noise signals and correlated pulses at amplitudes larger than a single photon-equivalent. This enables the use of SiPMs for triggering in low-amplitude environments and is crucial for the use of these sensors with high degrees of radiation damage and correspondingly large noise rates.

Overall, the limited radiation hardness of SiPMs presents one of the major challenges for the use of these devices in regions of high neutron fluence at accelerator-based experiments. The main macroscopic effect of radiation damage is the significant increase of dark current below and above breakdown, manifesting itself as an increase of the dark count rate. This also results in a loss of single photon resolution, which occurs for presently available SiPMs at a fluence of $10^{9}-10^{10}~\mathrm{n}_\mathrm{eq}/\mathrm{cm}^{2}$ when operated at room temperature \cite{Garutti:2018hfu}. The operation at low temperatures on the order of $-30$ to $-40~^\circ$C helps to reduce the dark rate, and with that also the damage-induced noise level, substantially. This enables the use of SiPMs also in areas of high fluence at the LHC for applications with typically large signals and moderate requirements on the signal-to-noise, such as calorimeters. More stringent requirements are imposed by tracking detectors, where the SiPMs are used in environments with more moderate radiation load. Detectors requiring single photon detection with high purity, such as imaging Cherenkov detectors, at present still remain the domain of other photon sensors. 

A general shortcoming remains the size of sensors, which is limited by the tolerable dark rate and the capacitance which influences the signal shape and the time resolution. This issue is addressed in practice either by using a larger number of electronics channels imposed by the smaller sensor size, or by combining the signals of multiple SiPMs arranged as an array via an active or passive analog circuit. In the latter case of passive ganging, the connection of several SiPMs in series provides the best performance, albeit at the prize of higher operation voltages. With such schemes, effective active sizes per electronics channel comparable to the use of photomultipliers and other photon detectors are achieved in present-day systems. 

The limited dynamic range imposed by the finite number of pixels of the sensors is dealt with in several ways by different experiments. One option is the use of sensors with smaller pixel sizes, which comes at the prize of a reduced photon detection efficiency. Larger sensor areas, which automatically provide a larger dynamic range, are viable solutions if the lower range of the relevant signal is at the few photon level with the larger detection surface and the increased noise rate does not result in a performance degradation. Sensors with very short recovery time, coupled with scintillators with an emission time longer than that recovery time, can provide an effective dynamic range that is substantially larger than the number of SiPM pixels and can thus extend the quasi-linear range of the SiPM response, but result in an increased sensitivity to afterpulsing. Finally, software correction procedures are used to linearize the response and to correct for saturation effects up to high signal levels. 

The sensitivity of the breakdown voltage, and with that the gain and photon detection efficiency, to temperature often requires precise control of the environmental parameters and / or an active correction for temperature changes. A number of different cooling solutions are used in practise, ranging from water cooling near room temperature to cooling with Peltier elements, two-phase evaporative systems or single-phase cooling loops to sub-zero temperatures. Alternatively, or in addition, to the direct cooling of the sensors, an automatic adjustment of the bias voltage based on measured temperature and known parameters of the temperature dependence can be used to stabilize the response. In principle such an approach is viable in cases where the SiPM itself does not constitute a significant heat source due to large leakage currents originating from substantial radiation damage, which would present the risk of a thermal runaway in case of an automatic increase of the applied voltage with increasing temperature. 

The vast majority of nuclear and particle physics applications of SiPMs are connected to the detection of scintillation light from organic and inorganic scintillators. A variety of different techniques have been developed for the optical coupling of the sensor to the scintillator and to wavelength-shifting fibers, depending on requirements of light collection efficiency, response uniformity and mechanical and budgetary constraints. A common option for detectors with a fiber-based collection of scintillation light are dedicated connectors which provide a well-defined gap between the polished end of a wavelength-shifting fiber and the photon sensor as well as a precise alignment of the fiber to the active area designed to maximize the number of detected photons. Light concentrators are used to collect the light from larger surfaces, such as matrices of scintillating fibers, to limit the overall needed sensor areas. In particular for systems with high channel count, coupling schemes amenable to automatization and mass production are being developed, such as the direct coupling of the photon sensor to scintillator elements making use of dome-shaped depressions optimized to achieve a uniform signal response over the full active area of the element, or the ``non-aligned'' coupling of SiPM arrays to end-polished scintillating fiber structures.

\begin{figure}[ht]
\centering
\includegraphics[width = 0.98\columnwidth]{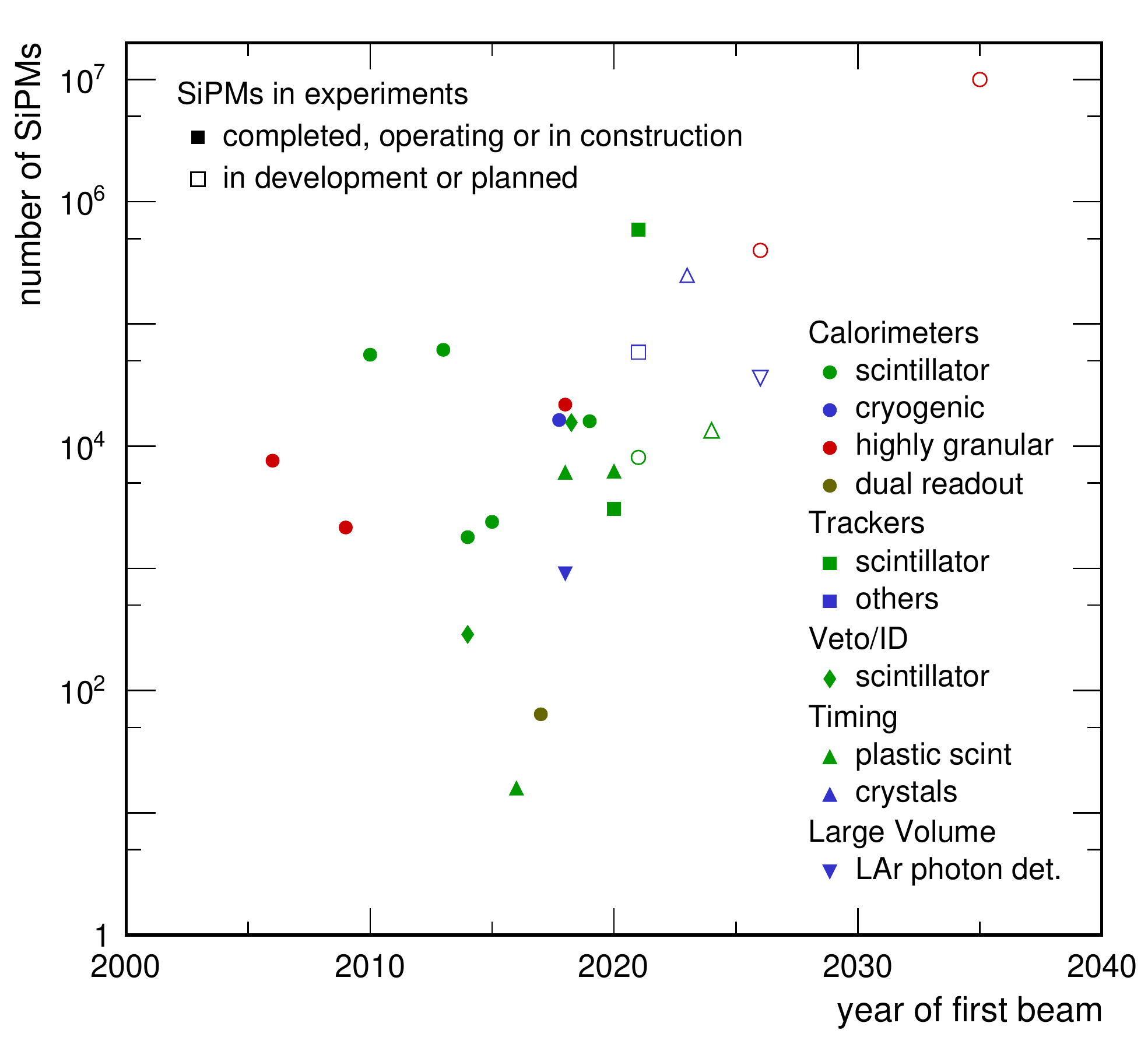}\\
\includegraphics[width = 0.98\columnwidth]{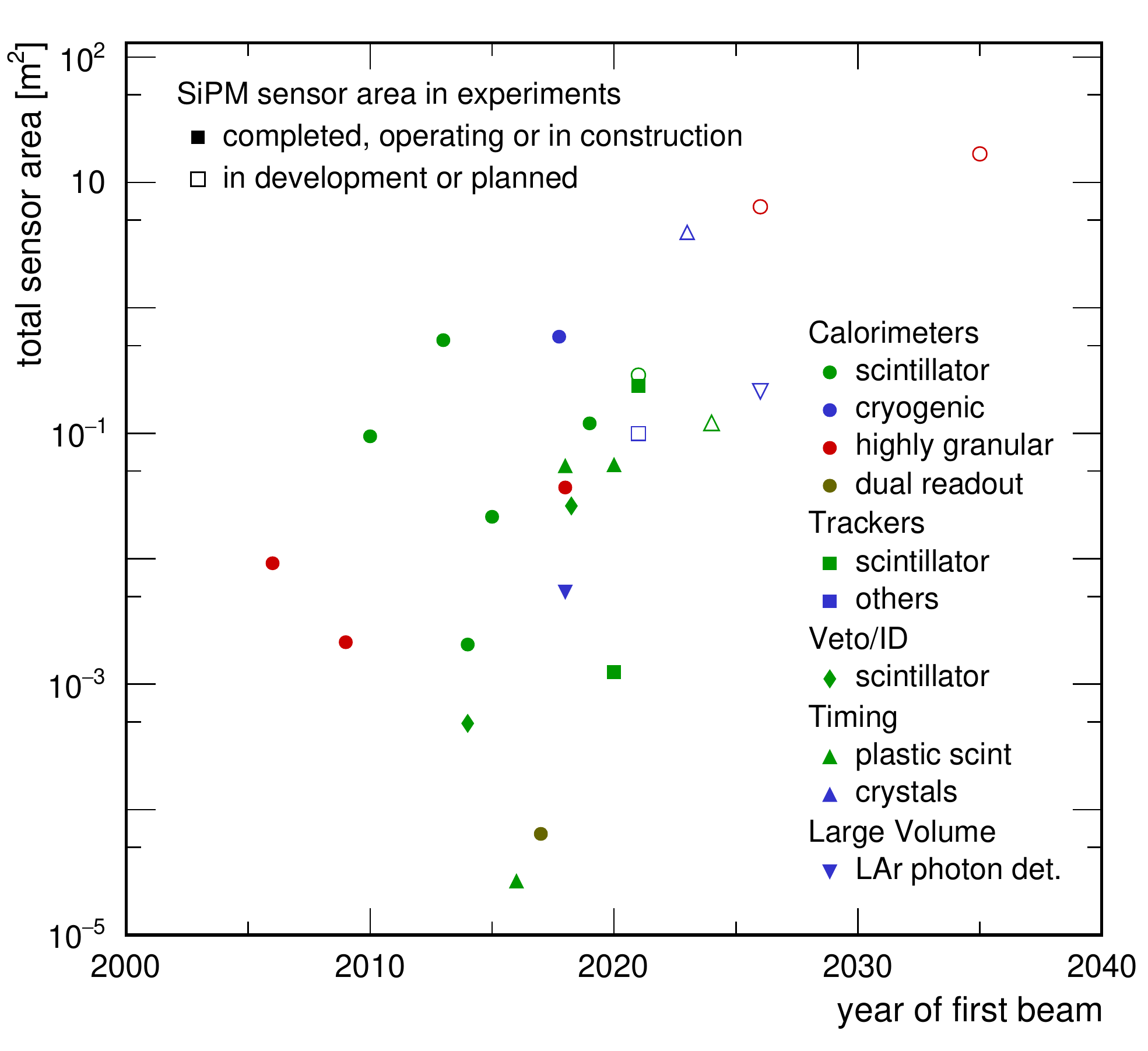}
\caption{Number of SiPMs (top) and the total SiPM sensor area (bottom) in detectors / experiments vs the year of first data taking discussed in this review. The T2K near detector is counted as a calorimeter, since the calorimeters account for the majority of the sensors. \label{fig:Trends:SiPMs}}
\end{figure}

\autoref{fig:Trends:SiPMs} shows the number of SiPMs as well as the approximate photon sensor area in the detector systems discussed in this review as a function of the year of the first beam for the system. For the number of SiPMs, individual units are counted also if they are electrically combined to reduce the number of readout channels or if several SiPMs sit on one common array. The division into projects under construction and those still in the R\&D or planning phase is arbitrary to a certain degree, with projects listed in the latter category if there are still substantial outstanding design choices. In particular in those cases the projected year of first beam still has quite some uncertainty. The assumptions made here are the interpretation of the author of currently available information. The figure illustrates the large variation of the size of systems using SiPMs, from small detectors using a few tens of SiPMs to those with more than 100,000, and plans extending to several millions.  It also shows that SiPM-based detectors in particle physics did not start ``small'', but with higher channel counts than comparable detectors with other photon sensors, well in line with the discussion of the reasons for the adoption of SiPMs above. The active sensor area of present systems is below 1~m$^2$, with several square meters expected for some next generation detectors. While this is very substantial, in particular when considering the small unit size, it is modest compared to the silicon area used in collider trackers and calorimeters, and also compared to the photocathode areas in large PMT-based detector systems. It is clearly apparent that the most common use for SiPMs in nuclear and particle physics, both in past and present as well as in future projects, is calorimetry, however with other applications, in particular timing detectors, gaining in relevance. 

The market of SiPMs is still evolving rapidly, with several competing producers, both fully commercial companies and collaborations of research institutes and industrial manufacturers. Currently the dominating producer for particle and nuclear physics applications is Hamamatsu, with the majority of the experiments and projects using their devices which are marketed under the name of Multi-Pixel Photon Counter (MPPC).  A wide variety of application-specific sensor types, in addition to standard ``off-the-shelf'' devices, are available,  addressing specific needs of experiments. These are often developed in a close collaboration of the research groups and the manufactures. Examples for such specific features are custom sensor sizes and packages, specialized protection layers and entrance windows to enable VUV sensitivity, and sensor arrays to provide a high degree of integration.

\section{Summary \& Conclusions}
\label{sec:Summary}

Silicon photomultipliers have seen a rapid adoption in particle and nuclear physics experiments in the last decade. This adoption is driven by properties and capabilities of SiPMs that address the needs of modern particle and nuclear physics experiments, among them insensitivity to magnetic fields, a high photon detection efficiency, low operating voltages and a very small size. Combined with the advances in modern microelectronics, which enable the construction of detector systems with large numbers of electronics channels, these features provide the foundation for new concepts in detector systems based on optical readout. SiPMs are replacing classical photomultipliers and other light detectors, but also enable new applications which were previously not possible, in particular in systems with very large numbers of electronic channels and / or stringent space constraints.  

Following first large scale application in an imaging hadronic calorimeter prototype by the CALICE collaboration from 2006 on, and the choice of SiPMs for the readout of all scintillator-based detectors in the T2K near detector ND280 in 2010, the increasingly wide-spread use of SiPMs in experiments in the last decade has brought this new technology to maturity. Today, SiPMs are used or planned to be used in almost all detector types with optical readout, with the SiPM-based options typically reaching a performance that is comparable to or surpassing the performance of similar systems with other light detectors. 

The design choices for the systems are driven by specific properties, limitations and requirements of SiPMs. Examples are the cold operation at $-30~^\circ$C or below in LHC detector systems to keep the radiation damage - induced dark rate at tolerable levels,  precise coupling units to match fibers to the active area of the sensor in cases where fibers are used to guide the light to the SiPM, software procedures to correct for the inherent non-linearity of the sensor response due to the finite number of pixels, and a precise control of the operating temperature, or an active stabilization of the bias voltage to eliminate the impact of the temperature dependence of the breakdown voltage. The possibilities of the sensors are exploited by specific shaping of scintillator elements to enable direct coupling of the photon sensor without the need of wavelength shifting fibers, the integration of multiple SiPMs with separate readout on a common die to provide highly granular coverage with minimal dead zones, and the detection of VUV light with sensors in specialized packages. 

With the ongoing downward trend of the costs of these devices and continuing improvements in performance and the uniformity of large samples, the adoption of SiPMs for light detection in nuclear and particle physics experiment is expected to continue, making them the standard solution in particular for scintillator-based detector systems.

\bibliographystyle{elsarticle-num}
\bibliography{SiPM_ParticlePhysics}

\end{document}